\documentclass[a4paper,10pt,accepted=2025-02-06]{quantumarticle}
\pdfoutput=1

\usepackage[utf8]{inputenc}
\usepackage[english]{babel}
\usepackage[numbers,sort&compress]{natbib}

\usepackage{graphicx}
\usepackage{amsmath, amssymb, amsthm}
\usepackage{tensor}
\usepackage{braket}
\usepackage{cases}
\usepackage{dsfont}
\usepackage[dvipsnames]{xcolor}
\usepackage{hyperref}
\usepackage{nameref}
\hypersetup{
   colorlinks = true,
   allcolors = RoyalPurple
}
\usepackage[percent]{overpic}
\usepackage{svg}

\usepackage{tikz}
\usetikzlibrary{decorations.pathmorphing}
\usetikzlibrary{decorations.pathreplacing}
\usetikzlibrary{decorations.text}
\usetikzlibrary{shapes}
\usepackage{quantikz}
\usepackage{pgfplots}
\pgfplotsset{compat=1.18}

\usepackage[export]{adjustbox}
\usepackage{leftidx}
\usepackage{BOONDOX-cal} % more calligraphic letters, in particular H
\usepackage{orcidlink}

\usepackage{algorithm}
\usepackage{algpseudocode}

\DeclareMathOperator{\Tr}{Tr}

\DeclareMathOperator{\Span}{span}

\newtheorem{definition}{Definition}[]

\newtheorem{theorem}[definition]{Proposition}

\newcommand{\norm}[1]{\left\lVert #1 \right\rVert}
\newcommand{\abs}[1]{\left\lvert #1 \right\rvert}

\renewcommand{\epsilon}{\varepsilon}

\renewcommand{\braket}[1]{\langle #1\rangle}
\renewcommand{\ket}[1]{\vert #1 \rangle}
\renewcommand{\bra}[1]{\langle #1 \vert}

\newfloat{subroutine}{tbh}{for}%[part]
\floatname{subroutine}{Subroutine}

\definecolor{natfak}{HTML}{228848}
\definecolor{dunkelblau}{HTML}{0061A0}
\definecolor{tfmetal}{HTML}{0b354d}

\newcommand*{\textcite}{\citet}

\begin{document}
\title{Efficient Quantum Cooling Algorithm for Fermionic Systems}

\author{Lucas Marti\orcidlink{0009-0005-3130-380X}}
\email[]{Lucas.Marti@fau.de}
\affiliation{Department of Physics, Friedrich-Alexander Universität Erlangen-Nürnberg (FAU), Staudtstraße 7, 91058 Erlangen}

\author{Refik Mansuroglu \orcidlink{0000-0001-7352-513X}}

\affiliation{Department of Physics, Friedrich-Alexander Universität Erlangen-Nürnberg (FAU), Staudtstraße 7, 91058 Erlangen}

\author{Michael J. Hartmann \orcidlink{0000-0002-8207-3806}}
\affiliation{Department of Physics, Friedrich-Alexander Universität Erlangen-Nürnberg (FAU), Staudtstraße 7, 91058 Erlangen}

\date{2025-02-06}

\begin{abstract}
    We present a cooling algorithm for ground state preparation of fermionic Hamiltonians. Our algorithm makes use of the Hamiltonian simulation of the considered system coupled to an ancillary fridge, which is regularly reset to its known ground state. We derive suitable interaction Hamiltonians that originate from ladder operators of the free theory and initiate resonant gaps between system and fridge. We further propose a spectroscopic scan to find the relevant eigenenergies of the system using energy measurements on the fridge. With these insights, we design a ground state cooling algorithm for fermionic systems that is efficient, i.e. its runtime is polynomial in the system size, as long as the initial state is prepared in a low-energy sector of polynomial size. We achieve the latter via a pseudo-adiabatic sweep from a parameter regime whose ground state can be easily prepared. 
    We estimate that our algorithm has a polynomial runtime for systems where the spectral gap decreases at most polynomially in system size, and is faster than the adiabatic algorithm for a large range of settings. We generalize the algorithm to prepare thermal states and demonstrate our findings on the Fermi-Hubbard model.
\end{abstract}

\maketitle

\section{Introduction}

\begin{figure*}[ht!]
    \newcommand{\thermprocscaling}{1.33}
    \centering
    \scalebox{\thermprocscaling}{%
    \begin{tikzpicture}    
        \begin{scope}[yshift=2.2cm]
        \draw[thin] (.5,.5) -- (0,.5) node[at start,right] {$\color{teal}\ket{E_0}$};
        \draw[thin] (.5,1) -- (0,1) node[at start,right] {$\ket{E_1}$};
        \draw[thin] (.5,2) -- (0,2) node[at start,right] {$\ket{E_2}$} node[above,midway,yshift=.1cm] {\textit{System}};
        \shade[ball color=red!50] (.25,2) circle [radius=.1];
        \shade[ball color=red!50] (.15,1) circle [radius=.1];
        \shade[ball color=red!50] (.35,1) circle [radius=.1];
        \shade[ball color=red!50] (.12,0.5) circle [radius=.1];
        \shade[ball color=red!50] (.25,0.5) circle [radius=.1];
        \shade[ball color=red!50] (.38,0.5) circle [radius=.1];
        \draw [decorate,decoration={brace,amplitude=4pt,mirror,raise=4pt},yshift=0pt](0,2.01) --(0,.49)  node [midway,xshift=-.7cm] {$\ket{\tilde{E}_0}$};
        \draw[thin] (2,.5) -- (1.5,0.5) node[at start,right] {$\ket{0}$};
        \draw[thin] (2,1) -- (1.5,1) node[at start,right] {$\ket{1}$};
        \node[] at (1.75,1.4) {\textit{Fridge}};
        \shade[ball color=blue!50] (1.75,.5) circle [radius=.1];
        
        \node[below,fill=white,draw] at (1,.1) {\scshape Section \ref{sec:algoricooli}};
        \end{scope}
        
        \draw[dotted] (-1.1,1.75) -- (2.8,1.75);
        \begin{scope}
            \node[align=center] at (1.3,1.3) {Pseudo-adiabatic \\ Sweep};
            \draw[solid,-Stealth, align=center, decorate, decoration={snake,amplitude=0.02cm,segment length=.1cm},black] (-.7,3.2) to[bend right=35] (1,.8);
            \draw[solid,xshift=-.4cm, align=center, decorate, decoration={snake,amplitude=0.02cm,segment length=.1cm,text along path,text align=center,text={Sweep}},black] (-.7,3.2) to[bend right=35] (1.,.8);
            \node[scale=.9] at (.9,.5) {$\ket{\hat{E}_0}=\sqrt{1-\epsilon}{\color{teal}\ket{E_0}}+\sqrt{\epsilon} \ket{\psi}$};
            \node[below,fill=white,draw] at (1,.1) {\scshape Section \ref{sec:goodinitial}};
        \end{scope}
        
        \draw[dashed] (-1.1,-.5) rectangle (2.8,5.1);
        \node[below,fill=gray!10] at (1,5.35) {Initial State};
    \end{tikzpicture}
    \begin{tikzpicture}
        %% Thermalization
        \draw[dotted] (0.1,-.5) rectangle (3.2,1.75);
        \node[fill=white] at (1.5,1.75) {Cooling};
        \node[below,fill=white,draw] at (1.6,.1) {\scshape Section \ref{sec:algoricooli}};
        \node[scale=1,fill=gray!10] at (1.6,1.2)  {Subroutine \ref{sub:coolingstep}};
        % \node[scale=.6,fill=gray!10] at (1.6,1.1) {Algorithm \ref{sub:stoctherm}};
        \node[align=center,scale=.85] at (1.6,.6) {\{${\color{natfak}V_{(a,b)}},{\color{dunkelblau}V_{(e,f)}}$,...\}};
        %% SPectroscopy
        \draw[dotted] (0.1,2) rectangle (3.2,5.1);
        \node[below,fill=white] at (1.5,5.35) {Spectroscopy};
        
        \draw[tfmetal,thick] plot[smooth] coordinates {(0.4,2.7) (1.4,2.9) (2.8,2.7)};
        \draw[dunkelblau,thick] plot[smooth] coordinates {(0.4,2.7) (1,2.9)  (1.1,3.6) (1.3,2.8) (2,2.8) (2.1,3) (2.2,2.8) (2.8,2.7)};
        \draw[natfak,thick] plot[smooth] coordinates {(0.4,2.7) (0.6,2.8) (0.7,3) (0.8,2.8) (1.3,2.8) (1.4,3.2) (1.5,2.8) (2,2.9)  (2.1,3.6) (2.3,2.8) (2.8,2.7)};
        
        \draw[red,densely dashed,thick] (1.1,2.7)--(1.1,3.6);
        \draw[red,densely dashed,thick] (2.1,2.7)--(2.1,3.6);
        \draw[red,densely dashed,thick] (0.7,2.7)--(0.7,3.15);
        \draw[red,densely dashed,thick] (1.4,2.7)--(1.4,3.25);
        
        \draw[-Stealth] (0.4,2.7) -- (0.4,3.6) node[at end,above] {$E_F$};
        \node[natfak,scale=.65] at  (2.1,3.75) {$\omega_{k}^{(a,b)}$};
        % \node[natfak,scale=.65] at  (1.45,3.4) {$\omega_1^{(a,b)}$};
        \node[dunkelblau,scale=.65] at  (1.1,3.75) {$\omega_j^{(e,f)}$};
        \node[scale=.8] at (1.7,4.15) {\{${\color{natfak}V_{(a,b)}},{\color{dunkelblau}V_{(e,f)}},{\color{tfmetal}V_{(m,n)}}$,...\}};
        \node[below,fill=white,draw] at (1.7,2.6) {\scshape Section \ref{sec:ancontrolcool}};
        \node[scale=1,fill=gray!10] at (1.55,4.6) {Algorithm \ref{alg:bigbraincooling}};
        % \node[scale=.6,fill=gray!10] at (1.6,4.5) {Algorithm \ref{alg:controlledthermal}};
    \end{tikzpicture}
    \begin{tikzpicture}
        %% GIBBS
        \draw[dashed] (-.9,-.5) rectangle (1.9,5.1);
        \draw[dotted] (-.9,.6) -- (1.9,.6);
        
        \node[below,fill=gray!10] at (.5,5.35) {Final State};
        \begin{scope}[xshift=.75cm,yshift=2cm]
       \draw[thin] (-.3,.5) -- (.55,.5) node[at start,left] {$\color{teal}\ket{E_0}$};
        \draw[thin] (-.3,1) -- (.55,1) node[at start,left] {$\ket{E_1}$};
        \draw[thin] (-.3,2) -- (.55,2) node[at start,left] {$\ket{E_2}$} node[above,midway,yshift=.2cm,xshift=-.4cm] {\textit{System}};
        \shade[ball color=red!50] (.40,.5) circle [radius=.1];
        \shade[ball color=red!50] (.29,.5) circle [radius=.1];
        \shade[ball color=red!50] (.18,.5) circle [radius=.1];
        \shade[ball color=red!50] (.07,.5) circle [radius=.1];
        \shade[ball color=red!50] (-.04,.5) circle [radius=.1];
        \shade[ball color=red!50] (-.15,.5) circle [radius=.1];
        \end{scope}
         \node[below,scale=.8,align=center] at (.6,1.7) {for ground states:};
        \node[below,fill=white,draw,align=center] at (.6,1.3) {\scshape Section \ref{sec:subspacecooling}};
        \node[below,fill=white,draw,align=center] at (.6,.1) {\scshape Section \ref{sec:prepthermstates}};
        \node[below,scale=.8,align=center] at (.6,.5) {for thermal states:};

    \end{tikzpicture}
     \begin{tikzpicture}[overlay]
         \draw[thick,gray,-Stealth] (-6.6,1.5) -- (-5.9,1.5);
         \draw[thick,gray,-Stealth] (-6.6,1.7) -- (-5.9,2.7);
         \draw[thick,gray,-Stealth] (-3.3,1.5) -- (-2.6,1.5);
         \draw[thick,gray,-Stealth] (-3.3,3)  to[bend left=65] node[midway,-,draw=black,thin,midway,fill=white,text=black,decorate,decoration={snake,amplitude=0.02cm,segment length=.1cm},scale=.85] {$\{V,\omega\}$} (-3.3,1.8) ;
     \end{tikzpicture}

    }
    
    \caption{Schematic of the subspace cooling algorithm. \textbf{Initial state}: the system is initialized in the easy-to-prepare ground state $\ket{\tilde{E}_0}$ of a Hamiltonian $\tilde{H}$. \textbf{Pseudo-adiabatic Sweep}: A sweep reminiscent of an adiabatic sweep though much quicker steers the state towards an imperfect approximation $\ket{\hat{E}_0}$ of the desired ground state $\ket{E_0}$. \textbf{Spectroscopy} then \textbf{Cooling}: from this low-lying state, we perform the ancilla controlled cooling algorithm to identify the array of relevant couplers and resonances $\mathcal{A}=\{V,\omega\}$ within a small enough low-energy sector of the spectrum, and use this information to cool the imperfect state prepared once again, performing, this time around, the cooling steps directly on resonant frequencies. \textbf{Final State}: We obtain the desired ground state; this can be extend to thermal states with large inverse temperature $\beta$, where they are far from the maximally-mixed state.} 
    \label{fig:subspacecoolingsketch}
\end{figure*}
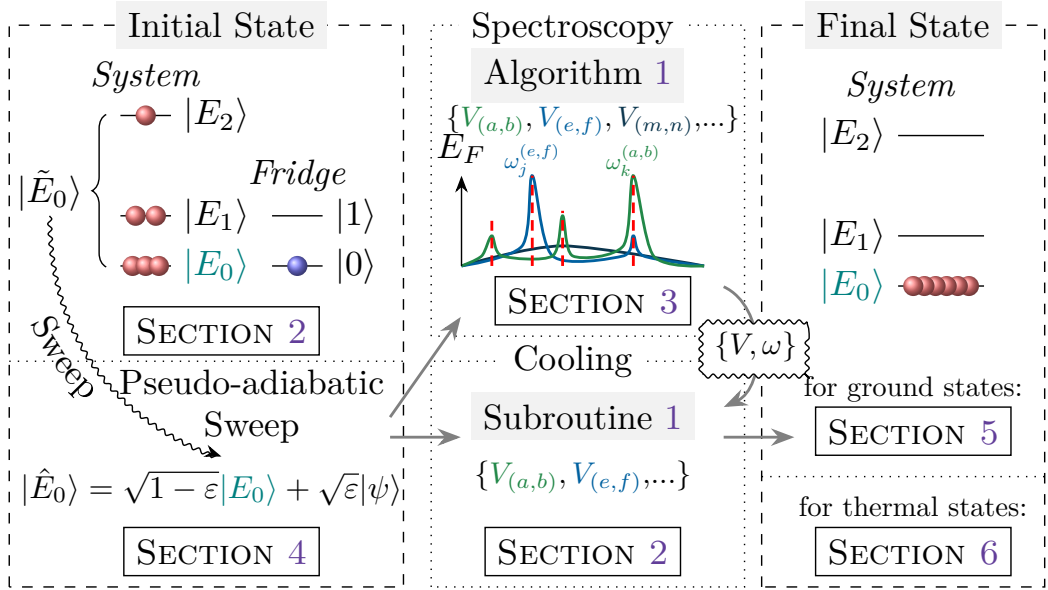

Ground states of quantum many-body Hamiltonians play a central role in condensed matter physics and often act as gateways to the solution of other classes of problems, such as optimization problems \cite{Farhi2022quantumapproximate}. Consequently, the search for algorithms that efficiently prepare them makes up a significant part of quantum algorithmic research. Due to the limitations of current or near-term noisy quantum (NISQ) devices, hybrid quantum-classical methods, such as variational quantum eigensolvers (VQE) \cite{Peruzzo_2014}, or generalizations thereof for Gibbs states \cite{chowdhury2020variational, Zoufal_2021}, have shown potential despite the shortfalls of imperfect hardware \cite{Moll_2018, Cerezo2021}. 

However, flaws that might not be systematically surmountable \cite{Bittel_2021} have steered research into new, different directions. Variational quantum algorithms suffer from sampling challenges \cite{Huang_2020, Zhenyu2020}, effects of noise on cost function evaluations \cite{Stilck_Fran_a_2021} and barren plateaus during training \cite{McClean_2018, Cerezo_2021, Wang_2021}. The situation is especially challenging for electronic structure Hamiltonians. Yet, these possess tremendous potential for applications from chemistry \cite{Aspuru_Guzik_2005, Arute2020, Wecker_2015} to strongly-correlated electrons \cite{Arovas_2021,Jiang_2018}, and have therefore received significant attention in the context of quantum simulation. Compounding the problems mentioned above -- as discussed by \textcite{Clinton_Sheridan_2024} -- the encoding of fermions to qubits is intrinsically non-local. Consequently, while quantum chips with more than 100 qubits \cite{Kim2023} have been developed, variational algorithms for fermionic systems have not been demonstrated on more than 12 qubits yet \cite{Zhao_Goings_2023}.

A switch to algorithms inspired by quantum many-body physics thus seems inevitable in order to circumvent the uncertainties of variational quantum computing. The price to pay is a scaling dependent on the physical properties of the studied Hamiltonian. Adiabatic quantum computing and its dependence on the energy gap of the time-dependent system of interest, for instance, was thoroughly investigated in this context \cite{Albash_2018}. 

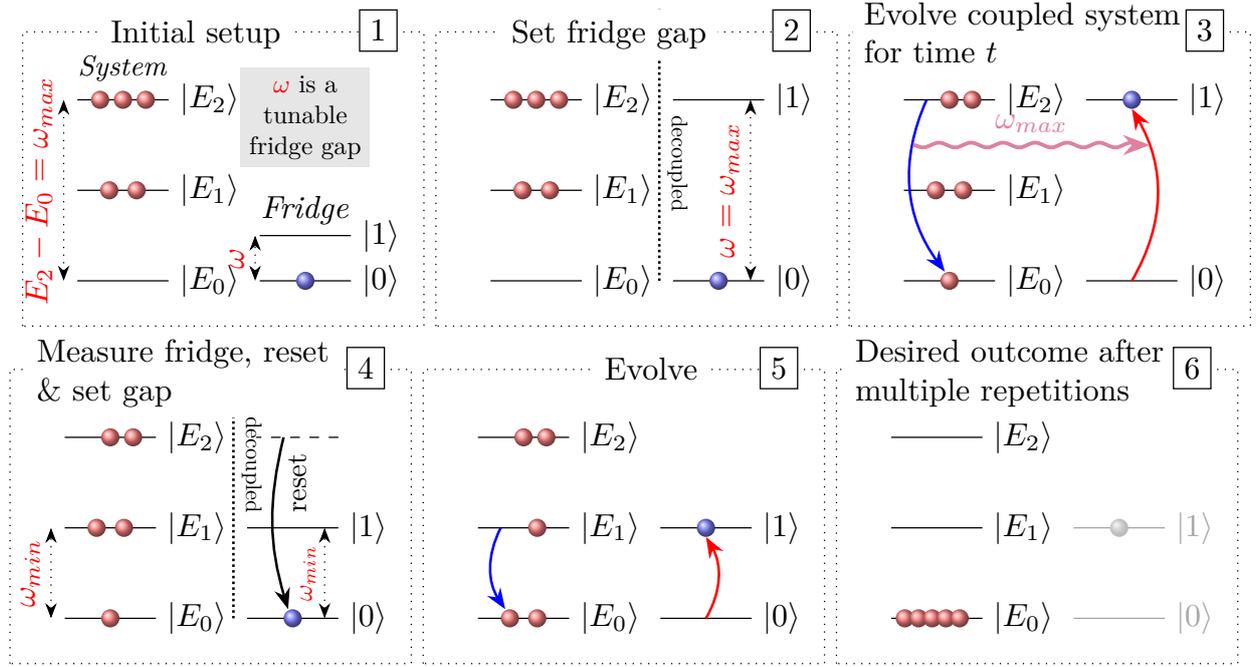
\begin{figure*}[ht!]
    \begin{center}
    \centering
    \newcommand{\coolingfigscaling}{1.2}
    \scalebox{\coolingfigscaling}{%
    \begin{tikzpicture}
        \draw[thin] (1,0) -- (0,0) node[at start,right] {$\ket{E_0}$};
        \draw[thin] (1,1) -- (0,1) node[at start,right] {$\ket{E_1}$};
        \draw[thin] (1,2) -- (0,2) node[at start,right] {$\ket{E_2}$} node[above,midway,yshift=.1cm,scale=.9] {\textit{System}};
        \shade[ball color=red!50] (.35,1) circle [radius=.1];
        \shade[ball color=red!50] (.65,1) circle [radius=.1];

        \shade[ball color=red!50] (.25,2) circle [radius=.1];
        \shade[ball color=red!50] (.50,2) circle [radius=.1];
        \shade[ball color=red!50] (.75,2) circle [radius=.1];
        
        \draw[thin,dotted,Stealth-Stealth,xshift=-.15cm] (0,0) -- (0,2) node[red,midway,above,rotate=90] {$E_{2}-E_{0}=\omega_{max}$};

        \draw[thin] (3,0) -- (2,0) node[at start,right] {$\ket{0}$};
        \draw[thin] (3,.5) -- (2,.5) node[at start,right] {$\ket{1}$} node[above,midway] {\textit{Fridge}};

        \draw[thin,dotted,Stealth-Stealth,xshift=-.05cm] (2,0) -- (2,.5) node[red,midway,above,rotate=90] {$\omega$};

        \node[above,align=center,scale=.8,fill=gray!20] at (2.5,1.25) {{\color{red}$\omega$} is a \\ tunable \\ fridge gap};

        \shade[ball color=blue!50] (2.5,0) circle [radius=.1];

        \draw[dotted] (-.6,-.5) rectangle (3.8,2.75);
        \node[below,fill=white] at (1.3,3) {Initial setup};
        \node[below,fill=white,draw] at (3.3,3) {\hypertarget{tikzcool1}{1}};
    \end{tikzpicture}
    \begin{tikzpicture}
        \draw[thin] (1,0) -- (0,0) node[at start,right] {$\ket{E_0}$};
        \draw[thin] (1,1) -- (0,1) node[at start,right] {$\ket{E_1}$};
        \draw[thin] (1,2) -- (0,2) node[at start,right] {$\ket{E_2}$};
        \shade[ball color=red!50] (.35,1) circle [radius=.1];
        \shade[ball color=red!50] (.65,1) circle [radius=.1];

         \draw[thick,densely dotted] (1.85,3) -- (1.85,0) node[yshift=-.2cm,midway,above,rotate=-90,scale=.7] {decoupled};

        \shade[ball color=red!50] (.25,2) circle [radius=.1];
        \shade[ball color=red!50] (.50,2) circle [radius=.1];
        \shade[ball color=red!50] (.75,2) circle [radius=.1];

        \draw[thin] (3,0) -- (2,0) node[at start,right] {$\ket{0}$};
        \draw[thin] (3,2) -- (2,2) node[at start,right] {$\ket{1}$};
        
        \draw[thin,dotted,Stealth-Stealth,xshift=-.15cm] (3,0) -- (3,2) node[red,midway,above,rotate=90] {$\omega=\omega_{max}$};

        \shade[ball color=blue!50] (2.5,0) circle [radius=.1];

        \draw[dotted] (-.6,-.5) rectangle (3.8,2.75);
        \node[below,fill=white] at (1.3,3) {Set fridge gap};
        \node[below,fill=white,draw] at (3.3,3) {\hypertarget{tikzcool2}{2}};
    \end{tikzpicture}
        \begin{tikzpicture}
        \draw[thin] (1,0) -- (0,0) node[at start,right] {$\ket{E_0}$};
        \draw[thin] (1,1) -- (0,1) node[at start,right] {$\ket{E_1}$};
        \draw[thin] (1,2) -- (0,2) node[at start,right] {$\ket{E_2}$};
        \shade[ball color=red!50] (.35,1) circle [radius=.1];
        \shade[ball color=red!50] (.65,1) circle [radius=.1];

        \shade[ball color=red!50] (.50,2) circle [radius=.1];
        \shade[ball color=red!50] (.75,2) circle [radius=.1];

        \shade[ball color=red!50] (.5,0) circle [radius=.1];

        \draw[thin] (3,0) -- (2,0) node[at start,right] {$\ket{0}$};
        \draw[thin] (3,2) -- (2,2) node[at start,right] {$\ket{1}$};

        \shade[ball color=blue!50] (2.5,2) circle [radius=.1];
        
        \draw[thick,blue,-Stealth] (0.25,2) to[bend right=30] (0.45,0.1);
        \draw[thick,red,-Stealth] (2.5,0) to[bend right=30] (2.5,1.9);

        \draw[dotted] (-.6,-.5) rectangle (3.8,2.75);

        \path[draw=purple!50, very thick, decorate,decoration={snake,amplitude=-.75},-Stealth] (.1,1.5) -- (2.7,1.5) node[midway,above,color=purple!50] {$\omega_{max}$};
        
        \node[below,fill=white,align=left] at (1.3,3.2) {Evolve coupled system \\ for time $t$};
        \node[below,fill=white,draw] at (3.3,3) {\hypertarget{tikzcool3}{3}};
        
    \end{tikzpicture}}
    \scalebox{\coolingfigscaling}{%
    \begin{tikzpicture}
        \draw[thin] (1,0) -- (0,0) node[at start,right] {$\ket{E_0}$};
        \draw[thin] (1,1) -- (0,1) node[at start,right] {$\ket{E_1}$};
        \draw[thin] (1,2) -- (0,2) node[at start,right] {$\ket{E_2}$};
        \shade[ball color=red!50] (.35,1) circle [radius=.1];
        \shade[ball color=red!50] (.65,1) circle [radius=.1];

        \shade[ball color=red!50] (.50,2) circle [radius=.1];
        \shade[ball color=red!50] (.75,2) circle [radius=.1];

        \shade[ball color=red!50] (.5,0) circle [radius=.1];

        \draw[thin] (3,0) -- (2,0) node[at start,right] {$\ket{0}$};
        \draw[thin] (3,1) -- (2,1) node[at start,right] {$\ket{1}$};
        \draw[thin,dashed] (3,2) -- (2.1,2);

        \shade[ball color=blue!50] (2.5,0) circle [radius=.1];

        \draw[thick,densely dotted] (1.85,3) -- (1.85,0) node[yshift=.2cm,midway,above,rotate=-90,scale=.7] {decoupled};

        \draw[thick,solid,-Stealth,xshift=-.15cm,align=center] (2.55,2) to[bend right=15] (2.6,.1) node[above,rotate=90,scale=.9,xshift=1.55cm,yshift=-.35cm] {reset};

        \draw[thin,dotted,Stealth-Stealth,xshift=-.15cm] (0,0) -- (0,1) node[red,midway,above,rotate=90] {$\omega_{min}$};
        \draw[thin,dotted,Stealth-Stealth,xshift=-.15cm,align=center] (3,0) -- (3,1) node[red,midway,above,rotate=90,scale=.8] {$\omega_{min}$};

        \draw[dotted] (-.6,-.5) rectangle (3.8,2.75);
        \node[below,fill=white,align=left] at (1.3,3.2) {Measure fridge, reset \\ \& set gap};
        \node[below,fill=white,draw] at (3.3,3) {\hypertarget{tikzcool4}{4}};
        
    \end{tikzpicture}
    \begin{tikzpicture}
        \draw[thin] (1,0) -- (0,0) node[at start,right] {$\ket{E_0}$};
        \draw[thin] (1,1) -- (0,1) node[at start,right] {$\ket{E_1}$};
        \draw[thin] (1,2) -- (0,2) node[at start,right] {$\ket{E_2}$};
        \shade[ball color=red!50] (.65,1) circle [radius=.1];

        \shade[ball color=red!50] (.50,2) circle [radius=.1];
        \shade[ball color=red!50] (.75,2) circle [radius=.1];

        \shade[ball color=red!50] (.35,0) circle [radius=.1];
        \shade[ball color=red!50] (.65,0) circle [radius=.1];

        \draw[thin] (3,0) -- (2,0) node[at start,right] {$\ket{0}$};
        \draw[thin] (3,1) -- (2,1) node[at start,right] {$\ket{1}$};

        \shade[ball color=blue!50] (2.5,1) circle [radius=.1];
        
        \draw[thick,blue,-Stealth] (0.25,1) to[bend right=30] (0.3,0.1);
        \draw[thick,red,-Stealth] (2.5,0) to[bend right=30] (2.5,.9);

        \draw[dotted] (-.6,-.5) rectangle (3.8,2.75);
        \node[below,fill=white] at (1.9,3) {Evolve};
        \node[below,fill=white,draw] at (3.3,3) {\hypertarget{tikzcool5}{5}};
        
    \end{tikzpicture}
    \begin{tikzpicture}
        \draw[thin] (1,0) -- (0,0) node[at start,right] {$\ket{E_0}$};
        \draw[thin] (1,1) -- (0,1) node[at start,right] {$\ket{E_1}$};
        \draw[thin] (1,2) -- (0,2) node[at start,right] {$\ket{E_2}$};
        
        \shade[ball color=red!50] (.15,0) circle [radius=.1];
        \shade[ball color=red!50] (.30,0) circle [radius=.1];
        \shade[ball color=red!50] (.45,0) circle [radius=.1];
        \shade[ball color=red!50] (.60,0) circle [radius=.1];
        \shade[ball color=red!50] (.75,0) circle [radius=.1];

        \draw[thin] (3,0) -- (2,0) node[at start,right] {$\ket{0}$};
        \draw[thin] (3,1) -- (2,1) node[at start,right] {$\ket{1}$};

        \shade[ball color=gray] (2.5,1) circle [radius=.1];

        \filldraw[white!20,opacity=.65] (2,-.5) rectangle (4,2.75);
        
        \draw[dotted] (-.6,-.5) rectangle (3.8,2.75);
        \node[below,fill=white,align=left] at (1.3,3.2) {Desired outcome after \\ multiple repetitions};
        \node[below,fill=white,draw] at (3.3,3) {\hypertarget{tikzcool6}{6}};
    \end{tikzpicture}}
    \end{center}
    
    \caption{Simplified schematic of the cooling process. \textbf{Initial setup}: We start with a system with Hamiltonian $H_S$ and a fridge with Hamiltonian $H_F$. The system has, in this example, three energy levels, and starts in an arbitrary state, which is to be cooled to the ground state (i.e. all population at $E_0$). The fridge only has two energy levels; its energy gap $\omega$ is tunable, and its own ground state is easily preparable. \textbf{Set fridge gap}: While system and fridge are decoupled, we initialize the fridge in its ground state and tune its gap to match $E_2-E_0=\omega_{max}$. \textbf{Evolve coupled system}: We couple the system to the fridge with a coupling term $V$ and evolve the entire ensemble with the Hamiltonian $H_S +\omega H_F + \alpha V$ for some time $t$. \textbf{Measure fridge}: The fridge is decoupled from the system, measured and reset to its ground state. The gap is then set to the smaller transition, which still needs to be cooled. \textbf{Evolve}: The evolution step is repeated to cool the second transition. \textbf{Desired outcome}: When the process has been repeated enough times, one expects the system to be in its ground state.}
    \label{fig:coolingprocess}
\end{figure*}

Another such algorithm originates from thermodynamic cooling \cite{lin2024thermodynamic, puente2024quantum} and ideas from quantum control. Here, a system of interest is coupled to an ancillary system, the fridge, which is initialized in its ground state. Local excitations can be transported from the system to the fridge \cite{matthies2023programmable} (see Fig.~\ref{fig:coolingprocess}) by simulation of the Hamiltonian
\begin{align}
  H = H_S \otimes \mathds{1}_F + \mathds{1}_S \otimes \omega H_F + \alpha V,
  \label{eq:coolham}
\end{align}
with a system Hamiltonian $H_S$ whose ground state is to be prepared, a two-level Hamiltonian $H_F$ describing the dynamics of the fridge and an interaction $V$ between the two. The fridge gap $\omega$ and the interaction strength $\alpha$ are tunable, but fixed during each evolution step. For suitable $\alpha$, regularly measuring the ancillary fridge after an appropriate evolution time, and then re-initializing it in its ground state extracts energy out of the system, steering it towards its ground state \cite{Polla_2021}, or a thermal state \cite{shtanko2023preparing}, if reset in a probabilistic manner \cite{Clark_2010}. Note the difference between this method and algorithmic cooling in the context of quantum control \cite{burgarth2008protocol, Scarani02, park2015heat, lin2024thermodynamic}. Recently, \textcite{mi2023stable} implemented this algorithm on a simple system. \textcite{kishony2023gauged} also studied a (non-local) topological de-excitation in this context. Past approaches, however, make use of knowledge about the shape of excitations within the spectrum, that is the shape of $V$ \cite{matthies2023programmable}, or rely on the success of heuristic energy scans \cite{Polla_2021}.

In this work, we introduce a model-agnostic approach to cooling that enables us to find resonant transitions between system and fridge without prior knowledge about the system's spectrum. To this end, we find appropriate interaction Hamiltonians $V$, which we call couplers for simplicity. The couplers with sufficient overlap with a swap of excitations, which is necessary for a resonant transition, are derived from a solvable theory that is connected to $H_S$ by a quench (for instance a non-interacting limit). The second requirement to fulfill the resonance condition is the tuning of the fridge gap $\omega$. We measure energy gaps of the spectrum of $H_S$ in a spectroscopic way, using information from the fridge temperature as a witness for resonant transitions. The fridge is considered to be a single, resettable qubit. We assume that methods to control the fridge are readily available and that the reset of the fridge, encoded into qubits, happens instantaneously and without error. This way, transition energies of $H_S$ can be captured without prior knowledge about the spectrum of the system.

As a first step towards electronic structure Hamiltonians, we focus on the Fermi-Hubbard model, which is a popular model in the field of quantum algorithms~\cite{Stanisic_mon_2021,Cade_2020} and poses significant challenges in the quantum information setting~\cite{ogorman_Electronic_2021a}. We also discuss due adaptations for fermionic systems, and address conceptual complications that arise from the fermionic nature of the problem, for instance maintaining the particle number conserving subspace within the total Hilbert space. 

We also address a fundamental limitation of cooling that originates in the initialization. A random initial state will lead to an exponential complexity in the worst case, since every energy transition needs to be cooled separately. We propose a possible approach to prepare an educated guess within a low-energy subspace using a pseudo-adiabatic sweep \cite{farhi2000quantum,Benseny_2021}. Given the polynomial size of the low-energy subspace, we restrict the algorithm to a polynomial number of interaction Hamiltonians that are engineered from the free-fermions model. This reduction makes the algorithm efficient. 

Our subspace cooling algorithm is summarized in Fig.~\ref{fig:subspacecoolingsketch}. After a pseudo-adiabatic sweep from the free model to the Hamiltonian of interest, we start the cooling routine starting from a low-energy subspace. In a first run, the transition energies are captured from our spectroscopy algorithm, which yields a list of transition energies for every coupler $V_k$ with respect to which a resonance is hit. These can then be repeatedly used for cooling. The advantage of obtaining the required energy gaps in a scan before the cooling lies in the fact that the requirements of spectroscopy and cooling are decoupled (see section \ref{sec:subspacecooling}). 

We further generalize the algorithm to the preparation of thermal states of finite inverse temperature $\beta$, which are known for their use in semi-definite programming \cite{Brandao17}. To this end, each coupler is drawn from a Boltzmann distribution parametrized by $\beta$ and the eigenenergies of the system. In the limit $\beta \to \infty$, we return to ground state preparation.

The subspace cooling algorithm is efficient and guaranteed to be faster than either sweep or cooling alone within a significant window in problem space, as we show in section \ref{sec:subspacecooling}. Moreover, cooling algorithms as ours are less limited by decoherence processes in the qubits than variational algorithms. For cooling algorithms to be successful, the cooling rate needs to exceed heating rates induced by decoherence, whereas for a variational approach, the entire algorithm needs to fit into the time window set by the coherence time of the qubits.

\section{Algorithmic Cooling for Fermionic Systems}\label{sec:algoricooli}

In this section, we review the Fermi-Hubbard model on which we develop the cooling algorithm. We recap the principal subroutine for a cooling step, which, for illustrative purposes, we describe assuming we have perfect access to all information in the quantum system, including its eigenbasis. We then generalize these results to a realistic case, in which our information is limited to what is classically tractable. In the general case, in which the ideal couplers are not known, we argue that, as long as the system and fridge are weakly coupled, the rotating wave approximation (RWA) holds, and the fundamental cooling mechanisms are functional. In fact, the cooling algorithm can be extended through the RWA to work for any set of couplers that admit constant overlap with the ideal couplers. We discuss this in Appendix \ref{app:rwa}, and we design such couplers for the Fermi-Hubbard model in section \ref{sec:freecouplers}.

\subsection{The Fermi-Hubbard model}

The Fermi-Hubbard Hamiltonian is an idealized model aiming to describe the major dynamics of electrons moving in a lattice potential generated by the atomic nuclei. One may obtain it from successive approximations applied to the full solid-state Hamiltonian, as is done, for example, in \textcite{Wecker_2015}. The resulting Hamiltonian defines a lattice with at most one up- and one down-spin fermion on each site. The two surviving dynamical terms account for the momentum of fermions and the repulsion between particles of opposite spin on one site, 
\begin{align}\label{eq:hamfh}
    H &= -t\sum_{\langle i,j\rangle,\sigma} \hat{a}^{\dagger}_{i,\sigma} \hat{a}_{j,\sigma} + U \sum_i \hat{a}^{\dagger}_{i,\uparrow} \hat{a}_{i,\uparrow}\hat{a}^{\dagger}_{j,\downarrow} \hat{a}_{j,\downarrow},
\end{align}
where $t$ is the hopping or tunneling rate, and $U$ the Coulomb interaction. The Fock operators $\hat a_i$ and $\hat a^\dagger_i$ annihilate or create, respectively, a fermionic mode on lattice site $i$.
This model, although a massive simplification from the full solid-state Hamiltonian, still draws interest as it contains the smallest amount of structure necessary to model strongly-correlated electrons \cite{Arovas_2021,Jiang_2018}.

The Fermi-Hubbard model follows the usual fermionic statistics. It differs from spin models by the characteristic partition of its Hilbert space. Indeed, for a given state, the number of electrons present in the state is conserved, meaning the Hamiltonian never creates or destroys electrons. This implies that one is interested in solving particular realizations of the Hamiltonian, and furthermore suggests that some spin configurations are more complex than others: for example, with a single fermion present, one effectively cancels the onsite term; the dynamics of the model become trivial. We thus consider the total number of up and down fermions $N_f$, as an important model parameter. 

There exists a supplementary symmetry since the Hamiltonian prohibits spin-flips. This means that every occupation eigenspace is further divided into spin sectors. In the present case, one must ensure, by choosing equivariant couplers, that the cooling process keeps the state in the correct fermionic occupation number. There are multiple ways to go about this task, though we will only present the one we found most practical; the remaining possibilities are presented in Appendix \ref{app:suppcouplers}.

We can straightforwardly compute the dimension $d$ of the restricted subspace. Since there are, for each spin sector, $k_{\uparrow,\downarrow}$ electrons for $n$ sites, one easily finds the number of allowed computational states. Disregarding degeneracies, this translates to, 
\begin{align}
    d = \begin{pmatrix} n \\ k_{\uparrow} \end{pmatrix}\begin{pmatrix} n \\ k_{\downarrow} \end{pmatrix} = \mathcal{O} \left( \prod_{j \in \{\uparrow,\downarrow\}}\frac{(n-k_j)^{k_j} n^{n}}{(n-k_j)^n {k_j}^{k_j}} \right),
\end{align}
via Stirling's approximation, where $j$ indicates that we are taking the product over each spin sector, which may have differing occupations. In the half-filling case, where $k_j=n/2$, this scaling is exponential: $d = \mathcal{O}(2^n)$.

Throughout this work, we use the Fock representation to develop our ideas and define our algorithms. Porting these to digital quantum computers requires converting operators to discrete, gate-based operations on qubits, which can be done by turning the fermionic operators into Pauli operators, using available encodings, such as Jordan-Wigner, Bravyi-Kitaev \cite{Whitfield_2017}, or some local encodings \cite{Whitfield_2019,Chien_Whitfield_2020}. In all our simulations, we use the Jordan-Wigner encoding for simplicity.

Our couplers are described in the Fock basis, and thus the full cooling Hamiltonian \eqref{eq:coolham} can be straightforwardly converted to a Pauli sum. One then uses a Trotter decomposition to discretize the time evolution \cite{Suzuki_85, Childs_2021}; one can also further compress gate sequences for a specific Hamiltonian \cite{Mansuroglu_2023, Mansuroglu_2023b}. A brief survey can be found in Appendix \ref{app:digitalizecooling}.

\subsection{The cooling step}\label{sec:naivecooling}
Ground state cooling consists of two repeated steps, Hamiltonian simulation and reset~\cite{mi2023stable,Polla_2021,puente2024quantum} (see Figure~\ref{fig:coolingprocess} for a sketch). In the former, we have to implement a time evolution generated by the Hamiltonian of Eq.~\eqref{eq:coolham}, in which we insert the Fermi-Hubbard Hamiltonian $H_S$ with eigenstates $\ket{E_k}$ for the system, and a one-qubit Hamiltonian $H_F = - Z/2 + \mathds{1}_F/2$ for the fridge, where $Z$ is a Pauli and $\mathds{1}_F$ the identity operator. $H_F$ has an easily preparable ground state $\ket{0}$. We then define a set of ideal couplers
\begin{align}
V_{(k,0)} := V_k &=  \ket{E_0} \bra{E_k} \otimes \ket{1} \bra{0} + H.c,
\label{eq:Vk}
\end{align}
in which each $k$ is associated to an eigenenergy $E_k$ and $\ket{0}$ is the ground state of $H_F$, while $\ket{1}$ is its excited eigenstate. We can show that the evolution under this Hamiltonian brings down population from the $k^\text{th}$ excited state to the ground state. 
\newpage
\begin{theorem}[informal]\label{thm:time}
    Let $H_k$ be the Hamiltonian that generates the dynamics for cooling the gap $E_k \rightarrow E_0$,
    \begin{align}
        H_k = H_S + \omega H_F + \alpha V_k,
    \end{align}
    where the fridge is resonant with the energy gap that is to be cooled. If the initial overlap with the state $\ket{E_k}$ is $q_k$, the time-evolved fridge energy reaches its maximum at $t=\frac{\pi}{2\alpha}$ with a peak value $\braket{H_F}_{\ket{\psi(t)}}= q_k^2$.
\end{theorem}
A detailed proof can be found in Appendix \ref{app:ancillacontrolledcooling}. These couplers swap excitations between system and fridge in a simulation time $t = \mathcal{O}\left( \pi/\alpha \right)$, but would require a full diagonalization of $H_S$ to be constructed. Note that the Hermitian conjugate in Eq.~\eqref{eq:Vk} also includes a heating term that raises the energy in the system. When applied to the fridge ground state, however, the term $\ket{0}\bra{1}$ vanishes, such that no heating is happening. This process is laid out in the inset below, which defines the elementary subroutine for cooling. 
\begin{subroutine}[H]
    \caption{Cooling step}\label{sub:coolingstep}
    \begin{algorithmic}[1]
        \Function{CoolingStep}{$\rho_S,\rho_F,V,\omega$}
        \State $\rho \gets \rho_S \otimes \rho_F$
        \State $\alpha \gets \omega/W, W \in \mathbb{R} \gg 1$
        \Comment{Fig.~\ref{fig:coolingprocess} panel \framebox{2}}
        
        \State $H \gets H_S + \omega H_F + \alpha V$
        \State Evolve for $t=\pi/\alpha$: $\rho$ $\gets$ $e^{i Ht}\rho e^{-i Ht}$
        \Comment{\framebox{3}}
        \State Measure fridge: $\rho_S, \braket{E_F} \gets \Tr_F( \rho ), \Tr(\rho E_F)$
        \Comment{\framebox{4}}
        \State\Return $\rho_S,\langle E_F \rangle$
    \EndFunction
    \end{algorithmic}
\end{subroutine}

\subsection{Fermionic couplers}\label{sec:freecouplers}
The ideal couplers $V_k$ require perfect knowledge of the target system, that is, its entire eigenbasis. This makes the entire pursuit futile. Choosing an arbitrary $V_k$ can, however, also result in an effective cooling of the system, provided it has non-vanishing overlap in operator space with the ideal couplers of Eq.~\eqref{eq:Vk}. Yet, the heating terms will not vanish, as opposed to the ideal case. We show in Appendix \ref{app:coolingderivation}, that these heating terms are suppressed, however, if the fridge starts in its ground state $\ket{0}$ and the interaction strength $\alpha$ is sufficiently small. In the following, we propose couplers that are close enough to the ideal couplers to efficiently cool the system.

Although it suffices to have a non-vanishing overlap to the ideal coupler, one still needs to make a reasoned choice in practice, and cannot pick, for example, random Pauli strings, whose overlap might be exponentially small in system size. In the context of fermionic systems, a random coupler will also not be equivariant with respect to particle number and spin symmetry. We derive a set of couplers that meet the desired properties, in a way that is specific to fermionic models, and which consists in making use of the related and analytically solvable non-interacting Hamiltonian. For the Fermi-Hubbard model, this means we set $U=0$ (or in a generic case, we discard any non-quadratic term), and we arrive at
\begin{align}\label{eq:hamfhnonint}
    \tilde{H} &= -t\sum_{\langle i,j\rangle,\sigma} \hat{a}^{\dagger}_{i,\sigma} \hat{a}_{j,\sigma},
\end{align}
also known as the free-fermions model. This Hamiltonian can be diagonalized via a Bogoliubov transform \cite{Jiang_2018} in tractable time $\mathcal{O}(N^3)$, since any quadratic Hamiltonian can, in one spin sector, be rewritten in $N \times N$ matrix form
\begin{align}
    \sum_{i,j} c_{i,j} \hat{a}^{\dagger}_{i} \hat{a}_{j} &= \left.\Vec{\hat{a}}^{\dagger}\right.^T M\Vec{\hat{a}} = U^{\dagger}  \left( \sum_i \epsilon_i  \hat{b}^{\dagger}_{i} \hat{b}_{i} \right) U,
\end{align}
where $M$ is the matrix of quadratic coefficients (in our case $c_{i,i+1}=c_{i+1,i}=-t$, i.e. the hopping terms), which is Hermitian by definition.  The matrix $M$ is diagonalized by the unitary $U$, which implements the Bogoliubov transform to the new Fock operators $\hat{b}_i$ and $\hat{b}_i^\dagger$, with eigenvalues $\epsilon_i$. The matrices $M$ and $U$ are both $N \times N$ and $\Vec{\hat{a}}$ is a length $N$ vector of Fock operators. We therefore have access to the energies of the free-fermions model and its eigenstates, the so-called Slater determinants. 

In a specific occupation subspace, Slater determinants can be created by using $N_f$ Bogoliubov operators $\hat{b}^{\dagger}$ on the vacuum, $\ket{\tilde{E}_j} = \hat{b}^{\dagger}_{n_{N_f}(j)} ... \hat{b}^{\dagger}_{n_0(j)} \ket{0}$. For simplicity, let us define the set of indices $I(j)$ such that $\prod_{n \in I(j)} \hat{b}^{\dagger}_{n} \ket{0}=\ket{\tilde{E}_j}$ and $\sum_{I(j)}\epsilon_j=\tilde{E}_j$. We use this to design more advanced couplers,

\begin{align}
    \ket{\tilde{E}_0}\bra{\tilde{E}_j} 
    &=\prod_{n\in I(0)} \hat{b}^{\dagger}_n\ket{0} \bra{0} \prod_{m\in I(j)} \hat{b}_m ,
\end{align} 
which we call the {\it free couplers}.
If we stay in the correct subspace at all times,
we may rewrite the coupler as $ \ket{\tilde{E}_0}\bra{\tilde{E}_j}  =\prod_{n\in I(0)} \hat{b}^{\dagger}_n \prod_{m\in I(j)}\hat{b}_m$,
from which we can recover the Fock operators in the original basis, by applying the inverse transform on the Bogoliubov operators. We can then obtain its Pauli form, for use in a gate-based algorithm, by applying a fermion-to-qubit encoding. The ground state of this non-interacting model is degenerate; there are therefore multiple ways to build the couplers. In practice, the choice among those doesn't impact their cooling efficiency; this also motivates our choice to neglect supplementary notation to keep track of degeneracies in the above expressions. If we need to identify the most relevant degenerate ground state, however, we can perturb the free states with a small Coulomb term. This lifts the degeneracy and we can identify which of the degenerate free ground states is closest to the interacting ground state.

It is natural to consider whether the free couplers have sufficient overlap with the ideal couplers, as this is a condition for cooling to be efficient. There are multiple hints suggesting this is the case: perturbation theory from the free states to the interacting states indicates that the free states make for very good $0^{\text{th}}$ order approximations. Furthermore, the interacting ground state is adiabatically connected to one of the degenerate ground states of the free model. Finally, we notice that other coupler designs, which seem to be also naturally fitting, fail to achieve any cooling. This includes Givens rotations and Coulomb couplers, which we also discuss in Appendix \ref{app:suppcouplers}.

One could also improve the runtime by skipping some couplers: since we also have the energies of the free-fermions spectrum at hand, we can associate a gap $\tilde{E}_j-\tilde{E}_0$ to each coupler $\ket{\tilde{E}_0}\bra{\tilde{E}_j}$, and, when cooling a specific energy gap, only use the couplers whose own energy gap is close enough. This avoids inefficient cooling steps, where we estimate the overlap between the free and exact eigenstates to be low, such as cases with large $|\tilde{E}_k - E_j|$. 

We may now use the free couplers as replacements for the ideal couplers. In general, the energy gaps of the interacting system will be different from the free gaps. Since an appropriate tuning of the fridge gap $\omega$ enables cooling transitions, we set our focus, in the following section, on spectroscopy.

\section{Ancilla Controlled Cooling}\label{sec:ancontrolcool}
To capture the energy gaps, one could na\"ively do a very fine linear scan of the entire spectrum as is done in \textcite{Polla_2021}, for instance. However, this approach means searching an exponentially large spectrum. We can improve on it by taking into account the information contained in the energy expectation value of the fridge. The energy $E_F$ of the fridge qubit, is defined as
\begin{align}
    E_F = \omega \braket{H_F}_{U_j (\ket{\psi_S} \otimes \ket{0_F})},
    \label{eq:EF}
\end{align}
where $U_j$ is the time evolution due to the application the coupler $V_j$ (see Eq.~\eqref{eq:Vk}); it tells us whether a transition has happened. The scan must be fine grained where needed, and fast around less important frequencies; we thus allow for a variable speed of the scan that can be controlled via $E_F$. That is, we dynamically adapt the scanning step size in order to hasten the scan far away from resonances, and slow it down in the opposite situation. One starts the algorithm with the tunable gap at a very large $\omega$, larger than the largest energy gap of $H_S$, and proceeds with a cooling step. If the fridge energy has not climbed, we are far from a resonance; no cooling is happening; we must move on: the control function outputs a large $\delta\omega$, the dynamical step size. As we move closer to a resonance, the fridge gets hotter, and the $\delta\omega$ smaller. At the end, when we reach a given threshold ideally around $E_1-E_0$, we interrupt the algorithm. In practice, this happens by setting an upper bound on the cooling step simulation time, or if the time budget permits, keep decreasing $\omega$ until the fridge no longer heats up. The minimal $\omega$ value then provides an upper bound for the populated eigenstates. 

Introducing a controlled scan through energy gaps not only optimizes the time budget, it also enables a spectroscopic analysis of the energy curve. For every coupler $V_{(a,b)}$ of some gap $E_b -E_a$, we can collect the energies $\omega^{(a,b)}_\mu, a > b$, for which the fridge energy has a peak, as those are energy gaps for which $V_{(a,b)}$ triggers a transition. The spectroscopic scan and cooling can be performed together in one single run or in multiple subsequent independent runs. The latter can have profound advantages in real noisy devices, as discussed further in section \ref{sec:subspacecooling}.

Pseudo-code describing the spectroscopic scan introduced above is provided in algorithm \ref{alg:bigbraincooling}.

\begin{algorithm}[H]
    \caption{Spectroscopy via ancilla controlled cooling}\label{alg:bigbraincooling}
    \begin{algorithmic}[1]
        \Function{Spectroscopy}{$\rho_S,\rho_F,\{V_{(j,k)}\}$}
        \State $\omega_0 = (1+\delta) \max(\left|E_k-E_0\right|), \delta \in \mathbb{R}^+$
         \State $\mathcal{A} \gets \{(V_{(j,k)},\{\varnothing\})\}$
         \Comment{$\mathcal{A}(V_{(j,k)}) = \{\omega^{(j,k)}_{\mu}\} $}
        \While{$\omega_n \geq E_1-E_0$}
        \ForAll{$V \in \{V_{(j,k)}\}$}
            \State $\rho_S,E_F\gets\Call{CoolingStep}{\rho_S,\rho_F,V,\omega_n}$
            \State Control function: $\delta\omega = f(\omega_n, E_F)$
            \State Get $\omega_{n+1} \gets \omega_n - \delta\omega$
            \If{ $V$ has a resonance at $\omega_n$}
                \State Add $\omega_n$ to $\mathcal{A}(V)$
            \EndIf
        \EndFor
        \EndWhile
        \State\Return $\rho_S,\mathcal{A}$
        \EndFunction
    \end{algorithmic}
\end{algorithm}

\begin{figure}[ht!]
    \centering
    \includegraphics[width=1\columnwidth]{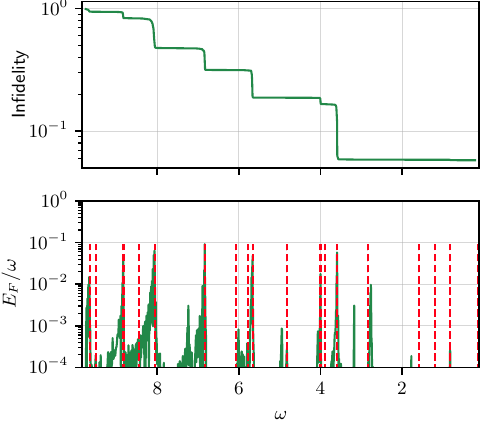}
    \caption{Controlled cooling spectroscopy on a 2 $\times$ 2 fermion lattice with the Fermi-Hubbard model at half-filling, that is, with 2 up-spin electrons, and 2 down-spin ones, with $t=1$, $U=2$. The system is initialized in a computational basis state of the qubit representation with the number of excited qubits equal to the number of considered fermions. The upper subplot shows the evolution of the infidelity of the cooled state with the ground state as the fridge gap $\omega$ is varied. The lower subplot shows the energy of the fridge qubit $E_F$ as a function of $\omega$. The red-dashed lines represent the true energy gaps of the system, within the 2-up and 2-down electron subspace. }
    \label{fig:normalcontrolledrun}
\end{figure}
We have defined the fridge to be a one-qubit system, with the single Pauli $Z$ Hamiltonian $H_F = - Z/2 + \mathds{1}_F/2$, in order to fix the eigenlevels to 0 and 1, and the corresponding eigenstates to $\ket{0}$ and $\ket{1}$. The fridge part of the coupler is the ladder operator $\ket{1}\bra{0}$. Note that, for all ensuing simulations, we use this same definition of the fridge, although any choice of fridge is possible, as long as the ground state is easily preparable, and resetting is unexpensive. Moreover, an increase in the number of fridge qubits can always be leveraged to parallelize the cooling of different energy gaps.

Throughout the work, we use \texttt{cirq} \cite{cirq_2023} and \texttt{openfermion} \cite{openfermion_2020} for noiseless statevector simulation (although noise is considered in Appendix \ref{app:noisestudy}). Fig.~\ref{fig:normalcontrolledrun} shows algorithm \ref{alg:bigbraincooling} in action. It uses the controlled function on a Fermi-Hubbard model, with the free couplers on the system. As the plot shows, we reach a fidelity of $0.942$ when the algorithm finishes.
Even if the energy gaps, represented by the dashed lines, were unknown, one could still identify them, due to the peaks in $E_F$; this is the expected spectroscopic quality of the controlled cooling algorithm. Not all the energy gaps are detectable, because the initial state may have no population in the respective levels, and there is therefore no excitation transfer. We start with a computational state in the fermionic subspace, and the scan reveals its eigenbasis amplitude decomposition. We discuss the control function used in the simulations in Appendix \ref{app:ancillacontrolledcooling}.

\section{Initial state preparation via a pseudo-adiabatic sweep
}\label{sec:goodinitial}

As discussed in the previous section, a swap of excitations results in heating the fridge qubit. We start in a state $\ket{\psi(0)}$ with population $P_j = |\braket{\psi(0) | E_j}|^2$ in the $j^\text{th}$ excited state that is to be cooled. If we used the ideal couplers $V_j = \ket{E_0} \bra{E_j} \otimes \ket{1} \bra{0} + H.c.$, the maximal energy of the fridge would be bounded by this initial population $P_j$ (see Appendix \ref{app:ancillacontrolledcooling}),
\begin{align}
    E_F \leq \omega P_j,
    \label{eq:temp_bound}
\end{align}
with $E_F$ as defined in Eq.~\eqref{eq:EF} (see also Proposition \ref{thm:time}). One intuitively understands that, when cooling a given gap, the maximal energy that can be extracted from it is proportional to the eigenstate population $P_j$. In the most na\"ive case, we start with a maximally mixed state, which has a population $P_j=d^{-1}$ that decays in the Hilbert space dimension $d$, rendering the probability to extract energy from the system exponentially small.

In practice, however, the coupler will not be ideal, but admits overlaps with multiple transitions, such that the upper bound in Eq.~\eqref{eq:temp_bound} is improved. Still, in the worst case, Eq.~\eqref{eq:temp_bound} gives a sample overhead of $1/P_j$. In order to keep the populations polynomial in system size, we must start in a state that concentrates the eigenstate population to a subspace of polynomial size. An additional benefit, although not necessary, comes when the subspace is low-lying in the energy spectrum, as this means populations are more often cooled to the ground state rather than to another excited state. 

\begin{figure}[ht!]
    \centering
    \includegraphics[width=1\columnwidth]{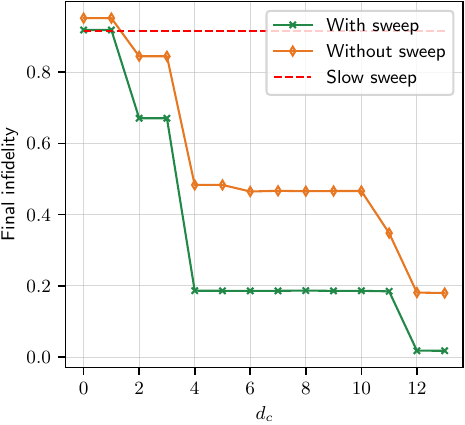}
    \caption{Improvement to infidelity from an initial pseudo-adiabatic sweep depending on the size of the cooled energy sector. $d_c$ indicates the number of cooled energy gaps, as well as the number of couplers used, and the y-axis shows the final infidelity of the controlled cooling spectroscopy algorithm over the spectrum up to the $d_c^{\text{th}}$ eigenlevel. The model is the same as in Fig.~\ref{fig:normalcontrolledrun}, although we start here in the ground state of the $t=0$ model, in which only the Coulomb interaction is present. The pseudo-adiabatic sweep accounts for less than $0.01\%$ of the total time budget, in only 5 Trotter steps. The horizontal line shows the minimal infidelity for a slow sweep, which is independent of $d_c$.}
    \label{fig:coulombfastsweepvm}
\end{figure}

Therefore, in this section, we introduce techniques to prepare states of low enough energy for use in cooling, by offloading the preparation to \textit{pseudo-adiabatic sweeps}. An \textit{adiabatic sweep} -- with which one eventually obtains the unique target ground state -- must run in a time that satisfies the adiabatic theorem~\cite{Albash_2018}, which can be very long. We therefore introduce the \textit{pseudo-adiabatic sweep}, whose sweep time is set to respect resource constraints rather than the time needed for the sweep to be adabatic. In some cases, however, no sweep, no matter how slow, can achieve a unity fidelity with the ground state, because of gap closings in the adiabatic passage or because of a degenerate initial ground state. For comparison, we still display them in Fig.~\ref{fig:coulombfastsweepvm} and \ref{fig:slaterfastsweepvm}. We call these sweeps \textit{slow sweeps}. In the absence of gap closings, this means that the final state is entirely contained within the degenerate ground state manifold; the slow sweep is then adiabatic in the sense that there is no leaking from one manifold to another, which conforms to theorem 1 from \textcite{Albash_2018}. When gap closings are present, the sweep is not adiabatic, and cannot be made so. In both cases, it is impossible to prepare the ground state, no matter how slow the sweep is. In this work, we use the distinction between \textit{adiabatic} and \textit{slow} to differentiate the case when a sweep can reach the actual ground state, and when it cannot, independently of its speed, respectively. Regardless of what causes the sweep to fail, in both figures, we have probed long enough sweep times to ascertain that the final fidelity has converged, and that, due to gap closings or degeneracy, we cannot prepare the ground state via a slow sweep.

We now use the pseudo-adiabatic sweep as a pre-processing step to the cooling algorithm. Again, by ``pseudo-'' we mean that the sweep time is dramatically reduced compared to adiabatic sweeps, such that, in practice, its new total runtime is a few orders of magnitude below the bound given by the adiabatic theorem \cite{Avron_Elgart_1999}. The system is then prepared, before the sweep, in an easy-to-prepare ground state. In the case of the Fermi-Hubbard model, this can be, for example, the Slater determinant (see section \ref{sec:freecouplers}) or the Coulomb state, the ground state of the opposite $t=0$ limit, which we use as initial state in Fig.~\ref{fig:coulombfastsweepvm}. Then, we steer this state towards an imperfect approximation of the desired ground state, so that we may then start cooling from there.

The effect of this preparatory step on the algorithm runtime is twofold: on the one hand, it brings the initial state closer to the target state, and, on the other, it shrinks the spectrum which one needs to cool, by suppressing the higher energy amplitudes. 

Fig.~\ref{fig:coulombfastsweepvm} shows the improvement in infidelity through the use of a pseudo-adiabatic sweep, compared to starting in the corresponding, non-swept, easy-to-prepare ground state. The graph shows that using an initial pseudo-adiabatic sweep may drastically reduce the spectrum to be scanned. The largest performance improvements thus emerge from only scanning a small sector of the eigenspectrum, where the pseudo-adiabatic sweep makes a crucial difference in the final result of the optimization, at almost no cost. A slow sweep has lower infidelity than the pseudo-adiabatic sweep, but the difference is barely discernible, due to the degeneracy of the initial ground states. We also display in Appendix \ref{app:suppcoolresults} the graph for the non-interacting or $U=0$ limit. In that case, the initial ground state is costlier to prepare \cite{Jiang_2018}, whereas the ground state of the Coulomb is a computational basis state, which can be prepared in constant depth.

\section{The Subspace Cooling Algorithm}
\label{sec:subspacecooling}

We now gather all the previously discussed topics to formally present the subspace cooling algorithm and study its efficiency.
Fig.~\ref{fig:subspacecoolingsketch} breaks down the different parts and indicates the sections in which they are introduced, hinting at the general structure of the subspace cooling algorithm. For completeness, we provide a full text description of the algorithm in this part, and further expand on the related theoretical scaling. In algorithm~\ref{alg:subspacecooling}, we show the pseudo-code version of the algorithm, and in algorithm~\ref{alg:thermaprotomain}, we show its generalization to thermal states.

We aim to find the ground state of a target Hamiltonian. The algorithm begins with the system in an easily preparable ground state $\ket{\tilde{E}_0}$. Then, we proceed with the pseudo-adiabatic sweep as described in section \ref{sec:goodinitial}, obtaining our first approximation of the target ground state $\ket{\hat{E}_0} = \sqrt{1-\epsilon} \ket{E_0} + \sqrt{\epsilon} \ket{\psi}$. Here, $\ket{\psi}$ is a superposition of excited states with coefficients decaying in energy difference \cite{Albash_2018} as long as the pseudo-adiabatic sweep time $t_{\text{ps}}$ is constant. From this low-energy state, we perform the controlled cooling algorithm in a small sector of the energy spectrum, using all the free couplers associated with gaps within this interval $V_{(j,0)}=\ket{\tilde{E}_0}\bra{\tilde{E}_k} \otimes \ket{1}\bra{0} + H.c.$. When the algorithm is done, spectroscopy of the couplers allows us to define sets of resonant gaps associated with each coupler, $V_{(j,0)}\leftrightarrow\{\omega_1^{(j,0)},\omega_2^{(j,k)},...\}$. If a coupler is found to have no particular resonances, it is discarded. The resulting list of correspondences is then passed on to the next step of the algorithm, during which we prepare a trial state with the pseudo-adiabatic sweep again, and may now use the information at hand to directly interact with resonant energy gaps. In the end, we obtain the desired ground state.

Although noiseless simulations enable us to perform the spectroscopy of all couplers in a single sequence, we expect that in a more realistic scenario, one obtains the information about couplers and their resonance through multiple applications of algorithm \ref{alg:bigbraincooling}, in which the inputs $\{V_{(j,k)}\}$ are subsets of all the desired couplers. In our simulations, we show the result of cooling via spectroscopy as well as the identified resonances in a single graph, because it effectively displays both sides of the subspace algorithm. Yet, we only need classical information from this process; the subspace algorithm does not require cooling to succeed during spectroscopy. For rather small systems, one may use hybrid spectroscopy techniques such as the one presented by \textcite{Chan_2023}.

We could design a naïve hybrid cooling algorithm by performing an adiabatic sweep on a system that is simultaneously coupled to the fridge. In an ideal case, this can be considered an adiabatic sweep on an open quantum system that undergoes a cooling channel. However, recent work has shown \cite{Wild_2016} that the adiabatic sweep is rendered inefficient by this coupling at any finite temperature of the fridge: this is why we considered a separated version that tackles the individual problems of the sweep and cooling in a more specific way.

Let us now look at the conditions needed for the subspace cooling algorithm to have an advantage over a full adiabatic sweep or a naïve cooling algorithm. Ideally, the goal of our algorithm is to try to reduce the number of gaps to cool. The desired outcome is an effective number of gaps with better scaling.
\newpage
\begin{theorem}\label{thm:efficiency}
    Let $H(s) = (1 - \lambda(s)) H_1 + \lambda(s) H_2$ generate a sweep that connects two arbitrary Hamiltonians $H_1$ and $H_2$ with a switch function $\lambda$ with $\lambda(0)=0$ and $\lambda(t_s)=1$ for a fixed total time $t_s$. Let $t_{\text{sub}}$ denote the time needed for ground state preparation using the subspace cooling algorithm with subspace dimension $d_c \in \mathbb{N}$, then a fixed target error $\epsilon = \frac{\alpha}{\Delta}$ can be achieved in times
    \begin{align}
        t_s &= \mathcal{O}\left( \frac{K}{\alpha \Delta^2} \right) \\
        t_{\text{sub}} &= \mathcal{O}\left( \frac{K \Delta}{\alpha \Delta_c^3} + \frac{d_c(d_c-1)}{\alpha} \right), \label{eq:tsubtime}
    \end{align}
     where $\Delta = \min_s (E_1(s)  - E_0(s))$ is the minimal gap, $K$ is some constant, and $\Delta_c = \max_j \min_s (E_{j}(s) - E_{j-1}(s)), \, j \leq d_c$ is the maximal energy difference between neighboring energy levels within the low-energy subspace $d_c$. 
    \begin{proof}
It is known \cite{Albash_2018} that the error $\epsilon$ of the sweep scales as $\epsilon = \mathcal{O}\left( \frac{K}{t_s \Delta^3} \right)$, with a constant $K = \mathcal{O}\left( \max_s \norm{ \partial_s H(s)}^2 \right)$, which is typically $\mathcal{O}(1)$. Note that, in degenerate cases, the minimal gap refers to the gap between the first two energy manifolds. The total time of the sweep is hence given by $ \mathcal{O}\left( \frac{K}{\epsilon \Delta^3} \right)$. Assume the error is $\epsilon = \frac{\alpha}{\Delta}$, as suggested by the rotating wave approximation in the worst case (see section \ref{app:rwa}), then $t_s = \mathcal{O}\left( {K}/(\alpha \Delta^2) \right).$

The smaller the gap, the larger the required time $t_s$ gets. We now show how cooling comes into play to mitigate this increase: instead of only using a slow sweep, we perform the pseudo-adiabatic sweep followed by cooling, whose combined time can be significantly shorter than $t_s$.
At the end of the pseudo-adiabatic sweep, if we only aim for an output state in the lower energy sector consisting of the $d_c$ lowest energy eigenstates, we can allow for an arbitrary error within the subspace. For the error to leak outside the low-energy subspace, we again use $\epsilon = \frac
{\alpha}{\Delta}$. The time for the sweep then becomes $t_{\text{ps}}=\mathcal{O}\left( \frac{K \Delta}{\alpha \Delta_c^3} \right)$.

The cooling routine, on the other hand, incorporates a quantum simulation with a time $t = \frac{\pi}{2\alpha}$ between resets. Assuming resets are instantaneous, the total time of the cooling routine is then just the number of cooled energy gaps times $t$. Since we are restricted to the lowest $d_c$ energy eigenstates, this number is $d_c-1$. Another factor $d_c$ comes into play that captures the mean number of repetitions to measure the fridge in the excited state (cf. Eq.~\eqref{eq:transfer}), assuming a maximally mixed state within the low-energy sector as a worst case. In total, we have $t_c = \mathcal{O}\left( \frac{d_c(d_c-1)}{\alpha} \right)$, and, summing this expression with the pseudo-adiabatic sweep time, we obtain Eq.~\eqref{eq:tsubtime}.

    \end{proof}
\end{theorem}
The bound for $t_{\text{sub}}$ becomes tighter, if one loosens the assumption that the output state of the pseudo-adiabatic sweep is maximally mixed among the low-energy subspace, but rather accounts for the loss of population outwards by considering the energy gaps $\Delta_j$ between the $j^\text{th}$ eigenenergy and the $d_c+1^\text{st}$ energy.

The combined algorithm is faster than any sweep for a fixed error threshold, if and only if the number of low-energy states is bounded. To get a feeling for this upper bound, we assume that both the adiabatic sweep and the combined routine exhaust the upper bounds for the total time that are derived above. In the case where the subspace cooling algorithm is faster, we have
\begin{align}
    t_{\text{sub}} \leq t_s \iff  \frac{K \Delta}{\alpha \Delta_c^3} + \frac{d_c(d_c-1)}{\alpha} - \frac{K}{\alpha \Delta^2} \leq 0
\end{align}

Assuming that $d_c(d_c-1) \approx d_c^2$, we get 

\begin{align}\label{eq:dcsquaredbound}
d_c^2 \leq \frac{K}{\Delta^2} \left( 1 - \frac{\Delta^3}{\Delta_c^3} \right).
\end{align}

This inequality needs to be consistently solved for $d_c$ and $\Delta_c$, which are dependent on another. It defines an allowed interval for $d_c$, for which the subspace cooling can be faster than the adiabatic sweep. Proposition \ref{thm:efficiency} considers an error which is dependent on the minimal gap. Both subspace cooling algorithms and adiabatic sweep then require this minimal gap to be polynomially far from zero in order to be efficient. Furthermore, the inequality in Eq.~\eqref{eq:dcsquaredbound} tells us that, given this polynomially decreasing minimal gap, there is a situation in which the subspace cooling algorithm is faster than the adiabatic sweep (provided both exhaust their upper bounds), while both algorithms run in a time growing polynomially with system size. 

In the case of the Fermi-Hubbard model, the spectral gap decreases, in the worst-case, polynomially: estimating the ground state energy of Hamiltonians with exponentially decreasing spectral gaps belongs to the PSPACE complexity class~\cite{fefferman_Complete_2018,deshpande_Importance_2022}, whereas it has been shown that for the Fermi-Hubbard Hamiltonian, it is QMA-complete. It is also fairly straightforward to numerically estimate this scaling~\cite{levy_solving_2022}. Hence, since we perform our sweep along a quench of the Fermi-Hubbard model, we know that, all along the passage, the minimal gap should also be decreasing polynomially. Therefore, we know that there are realizations of the Fermi-Hubbard model for which the subspace cooling algorithm runs in polynomial time and is faster than the adiabatic sweep.

Now, for small $\Delta$ and large enough $\Delta_c$, the contribution $\frac{\Delta}{\Delta_c}$ becomes negligible, such that the inequality approximately becomes 
\begin{align}\label{eq:dcgapbound}
d_c \leq \frac{\sqrt{K}}{\Delta}.
\end{align}

In our numerical experiments, the initial state of the sweep was degenerate: the right hand side of Eq.~\eqref{eq:dcgapbound} goes to infinity as the minimal gap closes. In this case, we notice that, as the sweep approaches a degeneracy, the approximation $\Delta \ll \Delta_c$ becomes justified, and the complete degenerate subspace is included in the low-energy subspace. Then, the subspace cooling algorithm will in any case be faster than the adiabatic sweep, as we can consider this initial degeneracy as additional error in the pseudo-adiabatic sweep, which can be corrected for in an appended cooling protocol. In such a case, we can still estimate if it is possible for the subspace cooling algorithm to run in polynomial time by estimating the number of populated levels of the subspace $d_c$. As long as the degenerate ground state-manifold contains a polynomial number of states, and assuming there are no exact gap closings along these levels, then $d_c$ should remain polynomial. This can be extended to gap closings: if only a polynomial number of levels is connected via gap closings to the polynomially-sized degenerate subspace, then, again $d_c$ remains polynomial. In these situations, the subspace cooling algorithm runs in polynomial time, and is faster than the adiabatic sweep, which itself has no bounded runtime.

We finally show a pseudo-code version of the subspace cooling algorithm, as it is graphically depicted on Figure~\ref{fig:subspacecoolingsketch}. The algorithm uses the {\sc PseudoSweep} subroutine defined in Subroutines \ref{sub:thermalfunctions}.

\begin{algorithm}[H]
    \caption{Subspace cooling}
    \label{alg:subspacecooling}
    \begin{algorithmic}[1]
        \Function{SubspaceCooling}{$\ket{\tilde{E}_0}, t_{\text{ps}}$,}
        \State $\rho_S \gets \Call{PseudoSweep}{\ket{\tilde{E}_0},t_{\text{ps}}}$
        \State $\rho,\mathcal{A} \gets \Call{Spectroscopy}{\rho_S,\{V_{(a,b)}\}}$
        \State $\rho_S \gets \Call{PseudoSweep}{\ket{\tilde{E}_0}, t_{\text{ps}}}$
        \ForAll{$\{(V_{(a,b)}, \{ \omega^{(a,b))}_\mu \}\} \in \mathcal{A}$}
        \State $\rho \gets\Call{CoolingStep}{\rho_S,\beta,V_{(a,b)},\omega^{(a,b))}_\mu}$
        \State $\rho_S \gets \Tr_F(\rho)$
        \EndFor
        \State\Return $\rho$
        \EndFunction
    \end{algorithmic}
\end{algorithm}

While proposition \ref{thm:efficiency} assumes a given set of couplers $\{V_{(j,0)}\}$, we comment on the necessary resources in order to find suitable couplers. One such way is spectroscopy, described in algorithm \ref{alg:bigbraincooling}. In a similar way as in proposition \ref{thm:efficiency}, we can derive a bound for the runtime.
\begin{theorem}\label{thm:spectroscopy}
    Consider the same setup as in Proposition \ref{thm:efficiency}. The time to run spectroscopy described in algorithm \ref{alg:bigbraincooling} is
    \begin{align}
        t_{\text{spec}} &= \mathcal{O}\left( \frac{K \Delta}{\alpha \Delta_c^3} + \frac{d_c(d_c-1)^2 \Delta_c}{\alpha \overline{\delta \omega}} \right),
    \end{align}
    with $\overline{\delta \omega}$ being the average step size in the spectroscopy protocol. 
\end{theorem}
The only difference to proposition \ref{thm:efficiency} is an additional factor $(d_c-1)$ and a factor $\frac{\Delta_c}{\overline{\delta \omega}}$. The former comes from the iteration over the set of free couplers for every transition, and the latter is an overhead factor depending on the granularity of the energy scan.

\section{Preparation of Thermal States}\label{sec:prepthermstates}
A slight generalization of the above cooling algorithm leads us to the concept of thermal (or Gibbs) states, 
\begin{align}
    \rho_S(\beta) = \frac{1}{Z_S(\beta)} e^{-\beta H_S}, \quad Z_S(\beta) = \Tr\left( e^{-\beta H_S} \right).
\end{align}
Thermal states are relevant in finite temperature settings, and are an important resource for quantum semi-definite programming \cite{Brandao17}. Therefore, preparing some of the non-trivial thermal states that admit large $\beta$ is of particular interest.

If we wish to prepare a non-trivial thermal state with inverse temperature $\beta_{target}$ from some initial state, we only need to consider a few changes to the original cooling process. First, instead of resetting the fridge to its ground state $\ket{0}$, we may reset it in a probabilistic manner according to $\beta_{target}$. Thus, after each measurement, we set the fridge to be in $\ket{0}\bra{0}$ with probability 
\begin{align}
    p_0 = \frac{\exp(\beta_{target} \omega/2)}{Z_F(\beta_{target})},
\end{align}
or $\ket{1}\bra{1}$ with probability $p_1 = 1-p_0$. The derivation of these probabilities are standard and restated in Appendix \ref{app:thermalstateeprep}. The second consideration one has to make regards the energy gaps: whereas, in a cooling situation, we are interested in pushing the entire population to a single eigenlevel, thermal states are distributed over all eigenstates. We ought then to consider all possible transitions from any single level to any other. 

Yet, thermal states may sometimes be easier to prepare by cooling than ground states. Thermal states with non-trivial values of $\beta$ have a population $P_j \propto \exp(-\beta E_j)$ in the $j^{\text{th}}$ eigenstate. After the pseudo-adiabatic sweep, we obtain an imperfect state whose populations are inversely proportional to their respective energy gaps and the sweep time \cite{Albash_2018}. Therefore, one has, after the sweep, a state that is very amenable to thermal states, even more so than to a ground state. This, however, does not happen when the starting ground state is degenerate: the population can spread among all initially degenerate eigenstates, leaving a distribution that can significantly differ from the desired thermal distribution. For instance in Fig.~\ref{fig:zipcoolthermal}, we find most of the population in the first excited state. If the degenerate subspace dimension is smaller than $d_c$, one can still successfully thermalize the state up to a satisfactory fidelity. 

In the case of the Fermi-Hubbard model, we can reach the desired thermal state with the free couplers, drawn from a probability distribution that reflects the statistics of the thermal state populations and is determined by the inverse temperature $\beta$ and the resonant energy gaps within the low-energy subspace (see Appendix \ref{app:circumventionexpscale}).

\subsection{Subspace Thermalization}

We display a pseudo-code version of the subspace thermalization algorithm discussed above, the counterpart to subspace cooling. The first inset, Subroutines \ref{sub:thermalfunctions}, defines utility functions for use in the algorithm, and, because of legibility concerns, we have decided to give them their own space. The first function describes the \textsc{PseudoSweep} that we have thoroughly used throughout this paper. These arguments are omitted in the pseudo-code, but, as for any sweep, one requires an initial Hamiltonian, and a target Hamiltonian. In the end, one obtains an approximation whose accuracy depends on the time spent in the sweep.

The second function covers the probabilistic reset. One picks a random float from a continuous uniform distribution and, if it is lower than an expression depending on $\beta$, returns the ground state of the one-qubit Hamiltonian, or else the excited state. This of course equivalent to a coin flip where the heads has a probability $\exp(\beta \omega/2)/Z_F(\beta)$. This allows us to get a fridge that correspond to the effective inverse temperature $\beta$ that we are targeting.

Finally, the third function describes the thermal step, the counterpart to the cooling step, defined in Subroutine \ref{sub:coolingstep}. The main difference between the two is the use of the \textsc{ProbabilisticReset}.

\begin{subroutine}[H]
    \floatname{subroutine}{Subroutines}
    \caption{Utility functions for thermalization and cooling}
    \label{sub:thermalfunctions}
    \begin{algorithmic}[1]
        \Function{PseudoSweep}{$\ket{\tilde{E}_0}, t_{\text{ps}}$}
            \State Evolve for $t_{\text{ps}}$: $\ket{\Phi}=U(t_{\text{ps}})\ket{\tilde{E}_0}$
            \State \Return $\ket{\Phi}\bra{\Phi}$
        \EndFunction
     \end{algorithmic}
      \begin{algorithmic}[1]
        \Function{Probabilistic reset}{$\beta$}
            \State Draw random $x \in [0,1]$
            \If{$x < \exp(\beta \omega/2)/Z_F(\beta)$}
            \State \Return $\ket{0}\bra{0}$
            \Else
            \State \Return $\ket{1}\bra{1}$
            \EndIf
        \EndFunction
    \end{algorithmic}
    \begin{algorithmic}[1]
        \Function{ThermStep}{$\rho_S,\rho_F,V,\omega.\beta$}
            \State $\rho_F \gets \Call{ProbabilisticReset}{\beta}$
            \State $\rho_S \gets\Call{CoolingStep}{\rho_S,\rho_F,V,\omega}$
        \State\Return $\rho_S$
        \EndFunction
    \end{algorithmic}
\end{subroutine}

In the function \textsc{PseudoSweep}, $U(t_{\text{ps}})$ is the unitary implementing a curve in operator space, between two Hamiltonians $\tilde{H}$ and $H$, with time $t_{\text{ps}}$, such that $U(t_{\text{ps}})\ket{\tilde{E}_0}=\ket{E_0} + \sum_i c_i \ket{E_i}$, where $\ket{\tilde{E_0}}$ is the ground state of $\tilde{H}$, and the $\ket{E_i}$ are the eigenstate of $H$. The coefficient $\abs{c_i}^2$ follows $\abs{c_i}^2 = \mathcal{O}((E_i-E_0)^{-3} (t_{\text{ps}})^{-1})$.

This second inset describes the adapted controlled cooling algorithm in the case of thermalization. It works quite similarly to the former program, with the notable differences being the use of the probabilistic reset.

\begin{algorithm}[H]
    \caption{Spectroscopy in the case of thermalization}
    \label{alg:controlledthermal}
    \begin{algorithmic}[1]
        \Function{SpectroscopyTherm}{$\rho_S,\{V_{(a,b)}\},\beta$}
         \State $\rho_F \gets$ \Call{Probabilistic reset}{$\beta$}
          \State $\rho$, $\mathcal{A} \gets \Call{Spectroscopy}{\rho_S,\rho_F,\{V_{(a,b)}\}}$
    \State\Return $\rho$, $\mathcal{A}$
    \EndFunction
    \end{algorithmic}
\end{algorithm}

The final lines of pseudo-code we present here correspond to the subspace cooling algorithm from Section \ref{sec:subspacecooling}, adapted for thermalization, in a version that fits within the pseudo-code style we have developed over the course of the work. Fig.~\ref{fig:subspacecoolingsketch} sketches the overall operation of the cooling case. The algorithm below gives a more structured and exact picture, while also being much more concise. We again very briefly summarize the program. After the trial state is created with a pseudo-adiabatic sweep, we perform a spectroscopy of the couplers in the low-energy sector of the spectrum, and create a list $\mathcal{A}$ of resonant couplers and their resonances. At the end, we use the thermal step to cool towards the target Gibbs state, using the information gained beforehand, in this case, the array $\mathcal{A}$. In Appendix \ref{app:circumventionexpscale}, we describe a stochastic thermalization step which improves on the thermal step, and can readily replace it in the overall process. The place it should take is indicated by the \textsc{(Stoch)} in the algorithm below.

\begin{algorithm}[H]
    \caption{Subspace thermalization}
    \label{alg:thermaprotomain}
    \begin{algorithmic}[1]
        \Function{SubspaceTherm}{$\ket{\tilde{E}_0}, t_{\text{ps}},\beta$,$N_{steps}$}
        \State $\rho_S \gets \Call{PseudoSweep}{\ket{\tilde{E}_0},t_{\text{ps}}}$
        \State $\rho,\mathcal{A} \gets \Call{SpectroscopyTherm}{\rho_S,\{V_{(a,b)}\},\beta}$
        \State $\rho_S \gets \Call{PseudoSweep}{\ket{\tilde{E}_0}, t_{\text{ps}}}$
        \For{$N_{steps}$} 
        \State $\rho \gets\Call{(Stoch)ThermStep}{\rho_S,\beta,\mathcal{A}}$  
        \State $\rho_S \gets \Tr_F(\rho)$
        \EndFor
        \State\Return $\rho$
        \EndFunction
    \end{algorithmic}
\end{algorithm}

\begin{figure}[ht!]
    \centering
        \includegraphics{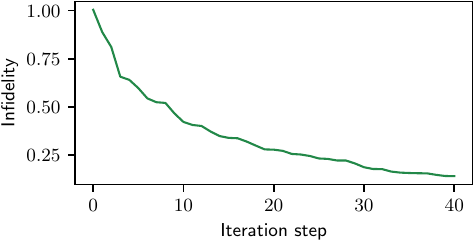}
        \caption{Thermalization using the thermal step in Subroutine \ref{sub:stoctherm}, with the free couplers, and information gained through a spectroscopy scan akin to that in Fig.~\ref{fig:normalcontrolledrun}. The model is the same as in that Figure; we however perform the pseudo-adiabatic sweep beforehand from the $t=0$ ground state, and from there target $\beta_{target}E_0=1$. In this case, we cool 4 energy gaps.} 
        \label{fig:zipcoolthermal}
\end{figure}

Fig.~\ref{fig:zipcoolthermal} shows that, using the information gained through spectroscopy, one succeeds in approaching a non-trivial target thermal state with $\beta_{target}E_0=1$ with a fidelity of $0.860$. In this particular case, we aim for a simulation with minimal resources and few iteration steps. Hence we use the energy gaps learned from the spectroscopy to keep the number of steps low, and for each step pick a different coupler, as mentioned above. Here, we have set $N_{steps}=40$.

\section{Outlook} 
In this work, we presented a hybrid algorithm for the ground state preparation of fermionic Hamiltonians. To this end, we combined the strength of quantum cooling routines with that of adiabatic sweeps to first learn about the spectrum of the system, and then systematically extract energy from the system. Our algorithm does not suffer from shortcomings known in variational algorithms, and it improves on the performance of its individual modules, an adiabatic preparation or a cooling routine starting with a random state.

While we did not explicitly port the subspace cooling algorithm to digital quantum computers, the discussion throughout this work points to a straightforward direction: the free couplers and the Hamiltonian are readily encoded with a fermion-to-qubit mapping, and the dynamics can be Trotterized. Experimental realizations should also be aided by the RWA: the free couplers do not need to be engineered exactly, but only approximately, as the weak coupling regime suppresses heating elements present in their approximate realization. Such a project seems to us quite promising as our work lays out the conceptual steps for its realization. 

We see several possibilities for further improvements. Throughout the work, we only used a single ancilla fridge for both our demonstrations and simulations. We foresee ways of reducing sample complexity by increasing the number of ancillas. A straightforward way would be assigning one ancilla for each coupler, which effectively parallelizes the cooling routine, but more sophisticated designs of the fridge are also conceivable.

There are also opportunities in the initial preparation of the state. Instead of using a pseudo-adiabatic sweep, one might consider either preparing a matrix product state on the device~\cite{malz_Preparation_2024}, or a perturbative solution of the Fermi-Hubbard model~\cite{li_Perturbation_2023}. 

Our simulations suggest that deviations from the RWA can be beneficial in situations where multiple transitions are cooled with a single coupler. In order to prove efficiency, however, we are limited to the RWA. A more adequate choice of couplers, possibly taking into account information about the form of excitations, or using perturbation theoretic treatment, is left for future research. 

\begin{acknowledgments}
    The authors thank Ignacio Cirac and Sirui Lu for inspiring discussions in an early stage of the project, as well as Tom O'Brien and Michael Lubasch for their constructive remarks on the manuscript. This work received support from the German Federal Ministry of Education and Research via the funding program quantum technologies - from basic research to the market under contract number 13N16067 “EQUAHUMO”. It is also part of the Munich Quantum Valley, which is supported by the Bavarian state government with funds from the Hightech Agenda Bayern Plus.
\end{acknowledgments}

\section*{Code Availability}
The code to reproduce the cooling simulations and all the figures in this work is available on GitHub~\cite{githubrepo}, at the following url: \url{https://github.com/lmarti-dev/fermionic_cooling}

\bibliographystyle{apsrev4-2}
\bibliography{sources}

\clearpage

\onecolumngrid
\appendix
\section{Derivation of the cooling algorithm}
\label{app:coolingderivation}

In this Appendix, we recap the basic functionality of the cooling routine. Our aim is to prepare the ground state of a problem Hamiltonian $H_S$. If the system is coupled to a fridge, described by a Hamiltonian $H_F$, via a coupler $V$, we can simulate the full dynamics governed by the Hamiltonian
\begin{align}
    H = H_S \otimes \mathds{1}_F + \mathds{1}_S \otimes \omega H_F + \alpha V, \qquad  H \in \mathcal{L}(\mathcal{H}) = \mathcal{L}(\mathcal{H}_S \otimes \mathcal{H}_F).
    \label{eq:appcoolham}
\end{align}
The constant $\alpha$ is a tunable coupling strength. We will consider system and fridge to be constructed from qubits. For the sake of simplicity, we discuss a fridge that consists of a single qubit and a system consisting of $n$ qubits.  A natural choice for the fridge Hamiltonian is thus $H_F = - \frac{1}{2}Z_F + \frac{1}{2} \mathds{1}_F$. If we were to cool a specific energy gap in the eigenspectrum, $\ket{E_j} \rightarrow \ket{E_k}$ with $E_j > E_k$, we can formally write down the coupler
\begin{align}
    V_{(j,k)} = \ket{E_k} \bra{E_j} \otimes \ket{1} \bra{0} + \ket{E_j} \bra{E_k} \otimes \ket{0} \bra{1}.
\end{align}
The second term is the hermitian conjugate of the first and ensures $V$ to be hermitian. If we start the fridge in its ground state $\ket{0}$, only the first term will have an effect on the system for short enough times. As we are interested in preparing the ground state of the system $\ket{E_0}$, the couplers that we aim for include the ladder operators $A_j := \ket{E_0} \bra{E_j}$. We could thus formally write down a suitable coupler $V$ for Eq.~\eqref{eq:appcoolham} 
\begin{align}
    V = \sum_j V_{(j,0)} = \sum_j \left( \ket{E_0} \bra{E_j} \otimes \ket{1} \bra{0} + \ket{E_j} \bra{E_0} \otimes \ket{0} \bra{1} \right).
    \label{eq:perfect_coupler}
\end{align}

\subsection{Rotating wave approximation}
\label{app:rwa}
\begin{figure}
    \scalebox{1}{
      \begin{tikzpicture}
        \draw[thin] (1,0) -- (0,0) node[at start,right] {$\ket{E_0}$};
        \draw[thin] (1,2) -- (0,2) node[at start,right] {$\ket{E_1}$};

        \shade[ball color=red!50] (.5,0) circle [radius=.1];

        \draw[thin] (3,0) -- (2,0) node[at start,right] {$\ket{\epsilon_0}$};
        \draw[thin] (3,2) -- (2,2) node[at start,right] {$\ket{\epsilon_1}$};

        \shade[ball color=blue!50] (2.5,2) circle [radius=.1];

        \draw[thick,blue,->] (0.25,2) to[bend right=30] (0.45,0.1);
        \draw[thick,red,->] (2.5,0) to[bend right=30] (2.5,1.9);

        \draw[dotted] (-.6,-.5) rectangle (3.8,2.75);

        \path[draw=purple!50, very thick, decorate,decoration={snake,amplitude=-.75},<->] (.1,1) -- (2.7,1) node[midway,above,color=purple!50] {$\alpha V$};

        \node[below,fill=white,align=left] at (1.9,3.2) {System cools \\ Fridge heats};
        \node[draw,thin,below,fill=white,align=left,scale=1.2] at (.5,3) {1};

      \end{tikzpicture}
      \begin{tikzpicture}
        \draw[thin] (1,0) -- (0,0) node[at start,right] {$\ket{E_0}$};
        \draw[thin] (1,2) -- (0,2) node[at start,right] {$\ket{E_1}$};

        \shade[ball color=red!50] (.5,2) circle [radius=.1];

        \draw[thin] (3,0) -- (2,0) node[at start,right] {$\ket{\epsilon_0}$};
        \draw[thin] (3,2) -- (2,2) node[at start,right] {$\ket{\epsilon_1}$};

        \shade[ball color=blue!50] (2.5,0) circle [radius=.1];

        \draw[thick,red,<-] (0.45,1.9) to[bend right=30] (0.45,0.1);
        \draw[thick,blue,<-] (2.55,0.1) to[bend right=30] (2.5,1.9);

        \draw[dotted] (-.6,-.5) rectangle (3.8,2.75);

        \path[draw=purple!50, very thick, decorate,decoration={snake,amplitude=-.75},<->] (.2,1) -- (2.7,1) node[midway,above,color=purple!50] {$\alpha V$};

        \node[below,fill=white,align=left] at (1.9,3.2) {System heats \\ Fridge cools};
        \node[draw,thin,below,fill=white,align=left,scale=1.2] at (.5,3) {2};

      \end{tikzpicture}
      \begin{tikzpicture}
        \draw[thin] (1,0) -- (0,0) node[at start,right] {$\ket{E_0}$};
        \draw[thin] (1,2) -- (0,2) node[at start,right] {$\ket{E_1}$};

        \shade[ball color=red!50] (.5,2) circle [radius=.1];

        \draw[thin] (3,0) -- (2,0) node[at start,right] {$\ket{\epsilon_0}$};
        \draw[thin] (3,2) -- (2,2) node[at start,right] {$\ket{\epsilon_1}$};

        \shade[ball color=blue!50] (2.5,2) circle [radius=.1];

        \draw[thick,red,<-] (0.45,1.9) to[bend right=30] (0.45,0.1);
        \draw[thick,red,->] (2.55,0.1) to[bend right=30] (2.5,1.9);

        \draw[dotted] (-.6,-.5) rectangle (3.8,2.75);

        \path[draw=purple!50, very thick, decorate,decoration={snake,amplitude=-.75},<->] (.2,1) -- (2.7,1) node[midway,above,color=purple!50] {$\alpha V$};

        \node[below,fill=white,align=left] at (1.9,3.2) {System heats \\ Fridge heats};
        \node[draw,thin,below,fill=white,align=left,scale=1.2] at (.5,3) {3};

      \end{tikzpicture}
      \begin{tikzpicture}
        \draw[thin] (1,0) -- (0,0) node[at start,right] {$\ket{E_0}$};
        \draw[thin] (1,2) -- (0,2) node[at start,right] {$\ket{E_1}$};

        \shade[ball color=red!50] (.5,0) circle [radius=.1];

        \draw[thin] (3,0) -- (2,0) node[at start,right] {$\ket{\epsilon_0}$};
        \draw[thin] (3,2) -- (2,2) node[at start,right] {$\ket{\epsilon_1}$};

        \shade[ball color=blue!50] (2.5,0) circle [radius=.1];

        \draw[thick,blue,->] (0.45,1.9) to[bend right=30] (0.45,0.1);
        \draw[thick,blue,<-] (2.55,0.1) to[bend right=30] (2.5,1.9);

        \draw[dotted] (-.6,-.5) rectangle (3.8,2.75);

        \path[draw=purple!50, very thick, decorate,decoration={snake,amplitude=-.75},<->] (.2,1) -- (2.7,1) node[midway,above,color=purple!50] {$\alpha V$};

        \node[below,fill=white,align=left] at (1.9,3.2) {System cools \\ Fridge cools};
        \node[draw,thin,below,fill=white,align=left,scale=1.2] at (.5,3) {4};

      \end{tikzpicture}
    }

    \caption{Possible actions of the coupler to the composite system. With the ideal couplers from Eq.~\eqref{eq:perfect_coupler}, as well as within the rotating wave approxmation ($\alpha \ll 1$), only option \textbf{1} happens.}
    \label{fig:altroute}
  \end{figure}
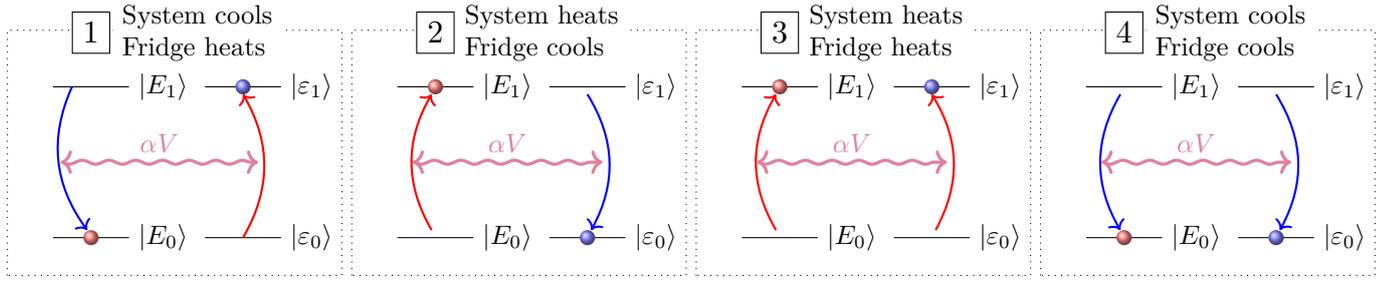

Obviously, we have no access to the eigenstates $\ket{E_j}$ a priori, which makes the coupler of the form of Eq.~\eqref{eq:perfect_coupler} unusable. If we instead assume no prior knowledge about the system, we have to start with a coupler that contains cooling terms as in Eq.~\eqref{eq:perfect_coupler}, but also heating terms and diagonal terms. We can always expand the coupling term in the energy eigenbasis, i.e.
\begin{align}
    V = \left[ \sum_{j<k} c_{jk} \ket{E_j} \bra{E_k} + h_{jk} \ket{E_k} \bra{E_j} + \sum_j d_j \ket{E_j} \bra{E_j} \right] \otimes \ket{1} \bra{0} + H.c.,
    \label{eq:imperfect_coupler}
\end{align}
where $H.c.$ stands for hermitian conjugate and the cooling and heating coefficients $c_{jk}, h_{jk}$ as well as the energy corrections $d_i$ are unknown, a priori. The cooling terms generate a swap of excitations between system and fridge, while the heating terms raise the energy of the system, effectively acting as a heating channel \cite{mansuroglu2024quantum} (see Fig.~\ref{fig:altroute}). We show that in the limit of weak coupling $\alpha \ll \omega$, all processes except for those that lead to cooling are  suppressed, as we can make use of the rotating wave approximation \cite{Wu07, Burgarth_2024} and assume $V$ to take the form of Eq.~\eqref{eq:perfect_coupler}. To do so, first consider Eq.~\eqref{eq:imperfect_coupler} in the interaction picture
\begin{align}
    \tilde V(t) &= e^{i t (H_S + \omega H_F)} V e^{-it (H_S + \omega H_F)} \nonumber \\
    &= \left[ \sum_{j<k} c_{jk} e^{it (\omega - \omega_{kj})} \ket{E_j} \bra{E_k} + h_{jk} e^{it (\omega + \omega_{kj})} \ket{E_k} \bra{E_j} + \sum_j d_j e^{it \omega} \ket{E_j} \bra{E_j} \right] \otimes \ket{1} \bra{0} + H.c.
    \label{eq:tilde_coupler}
\end{align}
where we defined $\omega_{kj} = E_k - E_j$ and assumed the eigenenergies to be ordered, i.e. $E_j \leq E_k$ if and only if $j \leq k$. Note that there is a remaining freedom of order within degenerate subspaces which is not of importance in the following. For the sake of clarity, we omit the identity operators in $H_S \otimes \mathds{1} \equiv H_S$ (similar for $H_F$) and clarify the support of the operators by the indices ${}_S$ and ${}_F$. Let us focus on the operator $\tilde V(t)$ that generates the dynamics in the interaction picture $\tilde \rho(t) = e^{it (H_S + \omega H_F)} \rho(t) e^{-it (H_S + \omega H_F)}$. We wish to approximate $\tilde V(t)$ by $\tilde V_{RWA}(t) = c_{jk} e^{it (\omega - \omega_{kj})} \ket{E_j} \bra{E_k} \otimes \ket{1} \bra{0} + H.c.$ that only contains one resonant transition, i.e. $\abs{\omega_{kj} - \omega} \ll \omega + \omega_{kj}$ for one fixed pair $k > j$. One can show (analogous to Lemma 1.2 in \textcite{Burgarth_2024}) that 
\begin{align}
    \norm{ \alpha \int_0^t (\tilde V(t) - \tilde V_{RWA}(t))} = \mathcal{O}\left( \frac{\alpha}{\omega - \omega_{ml}} \right),
    \label{eq:RWA_error}
\end{align}
for the operator norm $\norm{.}$ and with $(m,l) \neq (k,j) $ describing a non-resonant transition. The calculation of Eq.~\eqref{eq:RWA_error} for the Jaynes-Cummings model \cite{Burgarth_2024} can be straightforwardly applied to Hamiltonians in finite dimension. The key calculation is that the integrals over $e^{i \Omega s}$ are suppressed by a factor $\Omega$ for $t \geq \frac{\pi}{2 \alpha}$ making all oscillations negligible compared to the oscillation with frequency $\omega - \omega_{kj}$, if $\alpha \ll \omega - \omega_{ml}, (m,l) \neq (k,j)$. To be precise, the error contribution from the energy transition of the pair $(m,l)$ reads
\begin{align}
    \alpha \int_0^t ds \, c_{lm} e^{is (\omega - \omega_{ml})} \ket{E_l} \bra{E_m} \otimes \ket{1} \bra{0} + H.c.
    &= i c_{lm} \left( 1 - e^{it (\omega - \omega_{ml})} \right) \frac{\alpha}{\omega - \omega_{ml}} \ket{E_l} \bra{E_m} \otimes \ket{1} \bra{0} + H.c.,
    \label{eq:tilde_V_integrated}
\end{align}
and similar for the heating terms with coefficients $h_{lm}$ and the diagonal terms $d_{l}$. Using the fact that the error of the unitary time evolution ($U(t)$ generated by $\tilde V(t)$ and $U_{RWA}(t)$ generated by $\tilde V_{RWA}(t)$) is upper bounded by Eq.~\eqref{eq:RWA_error} \cite{Burgarth_2024, Mansuroglu_2023b}, $\norm{U(t) - U_{RWA}(t)} \leq \norm{ \int_0^t (\tilde V(t) - \tilde V_{RWA}(t))}$, yields a justification to approximating the time evolution by $\tilde V_{RWA}(t)$, or $V_{RWA}(t)$ being the corresponding operator in the Schrödinger picture. Note that the rotating wave approximation, while being broadly used in various quantum mechanical problems, has only recently been identified with a collection of approximation methods for Hamiltonian systems \cite{Venkatraman_2022, Burgarth2022oneboundtorulethem}.

In our setting, we start the fridge in the ground state, $\rho_0 = \rho_{S0} \otimes \ket{0} \bra{0}$, which ensures the cooling process $\ket{E_j} \bra{E_k} \otimes \ket{1} \bra{0}$ to be dominant. The above process has previously motivated other works \cite{Polla_2021, matthies2023programmable, mi2023stable, kishony2023gauged} to follow a simple sequence of cycles consisting of two steps, namely Hamiltonian simulation for $t \sim \frac{\pi}{2\alpha}$ and subsequent reset of the fridge to $\ket{0} \bra{0}$. 

\subsection{Theory of ancilla controlled cooling}
\label{app:ancillacontrolledcooling}
While the rotating wave approximation frees us from knowing the correct eigenstates before actually finding them, it relies on fine tuning of the parameters $\omega$ and $\alpha$, which are dependent on the eigenspectrum of the problem Hamiltonian. If we start with a maximally mixed state $\rho_{S0} = \frac{1}{d} \mathds{1}$, we need to cool all energy gaps; this has been implemented by a sweep through the energy spectrum \cite{Polla_2021}. This is of course wasteful, since it performs many non-cooling steps, and, depending on the step size $\delta\omega$, may miss some gaps. We discuss a novel, model-agnostic approach to solve this remaining problem by judiciously sweeping quickly through non-cooling intervals, and radically slowing down around resonances. We achieve this without disturbing the cooling process by measuring the energy of the fridge, a step that is already necessary for its reset. Note that multiple cooling steps and resets are needed for a single expectation value, the amount of samples depending on the overlap of the fridge qubit with its excited state. 

We now present a more formal version of Proposition~\ref{thm:time} formerly encountered in section \ref{sec:naivecooling}.

\setcounter{definition}{0}
\begin{theorem}[formal]\label{thm:formaltime}
    \label{theo:Hj_cool}
    Let $H_j$ be the Hamiltonian that generates the dynamics for cooling the gap $E_j \rightarrow E_0$, i.e.
    \begin{align}
        H_j = H_S + \omega H_F + \alpha c_{0j} \left( \ket{E_0} \bra{E_j} \otimes \ket{1} \bra{0} + H.c. \right),
    \end{align}
    where $H_F = - \frac{1}{2}Z_F + \frac{1}{2} \mathds{1}_F$ and the fridge is resonant with the energy gap that is to be cooled, i.e. $\omega = E_j - E_0 = \omega_{j0} =: \omega_j$. Let $\ket{\psi_S(0)}$ be a quantum state that admits overlaps $\braket{E_j | \psi_S(0)} = q_j$ with the $j^\text{th}$ excited state and $\braket{E_0 | \psi_S(0)} = q_0$ with the groundstate. The time-evolved fridge energy $E_F(t) = \omega \braket{H_F}_{U_j (\ket{\psi_S(0)} \otimes \ket{0_F})}$ reaches its maximum at $t=\frac{\pi}{2\alpha}$ with a peak value $\braket{H_F}_{\ket{\psi(t)}}= q_j^2$.
    \begin{proof}
        Without loss of generality and to make things easier to read in the following, we shift the energy spectrum to $E_0 = - \omega$. We can formally write down the four energy eigenstates of $H_j$ corresponding to the subspace $\mathcal{H}_j \cong \Span\{\ket{E_j}, \ket{E_0}\} \otimes \Span\{\ket{0}, \ket{1}\}$
    \begin{align}
        H_j \ket{\mathcal{E}_0} = - \omega \ket{\mathcal{E}_0} \qquad H_j \ket{\mathcal{E}_1} = \omega \ket{\mathcal{E}_1} \qquad &H_j \ket{\mathcal{E}_\pm} = \pm \alpha \ket{\mathcal{E}_\pm} \nonumber \\
        \ket{\mathcal{E}_0} = \ket{E_0} \otimes \ket{0} \qquad \ket{\mathcal{E}_1} = \ket{E_j} \otimes \ket{1} \qquad &\ket{\mathcal{E}_\pm} = \frac{1}{\sqrt{2}} \left( \ket{E_0} \otimes \ket{1} \pm \ket{E_j} \otimes \ket{0} \right)
        \label{eq:coupler_eigenstates}
    \end{align}
    All other energy eigenstates of $H_j$ stay product states of the form $\ket{E_k} \otimes \ket{0/1}_F$ and span the orthogonal complement $\mathcal{H}_1^\perp$, such that $\mathcal{H} \cong \mathcal{H}_j \oplus \mathcal{H}_j^\perp$. The initial state reads $\ket{\psi(0)} = \left( q_j \ket{E_j} + q_0 \ket{E_0} \right) \otimes \ket{0}_F + \ket{\perp}$ with $\ket{\perp} \in \mathcal{H}_j^\perp$. Since the dynamics is not mixing $\mathcal{H}_j$ and $\mathcal{H}_j^\perp$, we can focus on the component in $\mathcal{H}_j$. The initial state can be expanded into a superposition of energy eigenstates. The contributions in $\mathcal{H}_j$ read
    \begin{align}
        \left( q_j \ket{E_j} + q_0 \ket{E_0} \right) \otimes \ket{0}_F = q_0 \ket{\mathcal{E}_0} + \frac{q_j}{\sqrt{2}} (\ket{\mathcal{E}_+} - \ket{\mathcal{E}_-}) \quad \implies \quad \ket{\psi(t)} \Big\vert_{\mathcal{H}_j} = q_0 e^{i\omega t} \ket{\mathcal{E}_0} + \frac{q_j}{\sqrt{2}} (e^{-i \alpha t} \ket{\mathcal{E}_+} - e^{i\alpha t} \ket{\mathcal{E}_-}).
    \end{align}
    The operator $H_F$ has a simple action on the energy eigenstates, that is
    \begin{align}
        H_F \ket{\mathcal{E}_0} = 0 \qquad H_F \ket{\mathcal{E}_1} = \ket{\mathcal{E}_1} \qquad H_F \ket{\mathcal{E}_\pm} = \frac{1}{\sqrt{2}} \ket{E_0} \otimes \ket{1}_F = \frac{1}{2} \left( \ket{\mathcal{E}_+} + \ket{\mathcal{E}_-} \right),
    \end{align}
    which allows us to calculate expected occupation number of the fridge at time $t$,
    \begin{align}
        \braket{H_F}_{\ket{\psi(t)}} &= \frac{q_j^2}{4} \left( 1 - e^{2i \alpha t} \right) \left( 1 - e^{-2i \alpha t} \right) = \frac{q_j^2}{2} (1 - \cos(2 \alpha t)),
        \label{eq:transfer}
    \end{align}
    which starts at $\braket{H_F}_{\ket{\psi(0)}} = 0$ and has a peak at $\braket{H_F}_{\ket{\psi(t)}} = q_j^2$ at time $t = \frac{\pi}{2\alpha}$. 
    \end{proof}
\end{theorem}

The peak value is dependent on the initial state, in particular the initial occupations of $\ket{E_j}$ and $\ket{E_0}$. If we start with the pure high energy state, $\ket{E_j} \otimes \ket{0}$, then $\braket{H_F}$ can reach its maximal value 1. This dependence on initial population of $\ket{E_j}$ poses a problem for naive cooling. If we know nothing about the system, we must start close to the maximally mixed state for which the initial population and therefore the maximal excitation to swap is $\frac{1}{d}$, which can get exponentially small. To resolve this fundamental problem, we propose to start with an initial state that already lies in the low-energy sector. To this end, we perform a pseudo-adiabatic sweep as a pre-processor (see section \ref{sec:subspacecooling}). 

In the case where $\omega$ is off-resonant from any energy gap, the total interaction $V$ would be neglected in the weak coupling limit. The eigenstates of $H$ thus approximately decouple and the fridge state remains unchanged. So does $\braket{H_F}_{\ket{\psi(t)}}$. 

We take this as motivation to consider a measurement of $H_F$ as a witness for the resonance condition. If we assume no knowledge about the system for which we want to prepare the ground state, we will most probably start in a state that admits population throughout the whole energy eigenspectrum. For the sake of simplicity, suppose we start in a maximally mixed state $\rho(t=0) = \frac{1}{d} \mathds{1}$. The fridge energy can now be used to optimally control the speed of the sweep of $\omega$ throughout the eigenspectrum, namely deceleration at resonance and acceleration otherwise. This spectroscopic approach to algorithmic cooling is agnostic to model specific properties.

Starting in a value for $\omega$ that is larger than the spectral spread, we can iteratively and in a controlled way decrease $\omega$ after each step. Consider the $m^\text{th}$ cooling step in which we used a fridge gap $\omega^{(m)}$ and measured a final fridge energy $E_F^{(m)}$ (cf. Eq.~\eqref{eq:EF}). The next energy gap can be chosen as
\begin{align}
    \omega^{(m+1)} = \omega^{(m)} - f\left( E_F^{(m)} \right),
\end{align}
with a control function $f$ that decelerates when the fridge energy is maximal, $f(E_F^{(m)} = 1) = 0$, and accelerates whenever the fridge stays cool, $f(E_F^{(m)} = 0) = \Delta \omega_{max}$. There are a number of hyperparameters to be analyzed, amongst others the choice of $\Delta \omega_{max}$ and derivatives of $f$. Choosing $\Delta \omega_{max}$ too small (or $f'(1)$ too large) slows down the algorithm while a large value for $\Delta\omega_{max}$ (or a small value for $f'(1)$) leads to neglected energy gaps. Moreover, the position $T_c$ of the steepest point of $f$ controls the population that is transferred to the ground state. Similarly, a small $T_c$ slows down the process while a large $T_c$ tends to leave more population in the higher energy eigenstate. The control function used in the simulations is defined as 
\begin{align}
    f(E_F) &= x_1\exp\left(\frac{x_2}{\left(1-\log_{10}(E_F)\right)+x_3}\right)
    \label{eq:control_function}
\end{align}
It is largely motivated by phenomenological observations, as a compromise between simulation (wall) time and accuracy. There are three parameters. The first, $x_1$ control the overall scaling of the function, the second $x_2$ the sensitivity of the exponential, and the third $x_3$, the baseline speed at which the $\omega$ are swept.

\section{The role of couplers}
\label{app:suppcouplers}
In this section, we briefly discuss two supplementary methods to build couplers for the Fermi-Hubbard Hamiltonian, and we then present a decomposition of each coupler's individual role in the cooling process.
\paragraph{Symmetry-preserving couplers} 
One may look at the Pauli representation of the Fock operators, for example in the Jordan-Wigner encoding, from which one can extract terms such as $X_iX_{i+1} +Y_i Y_{i+1}$, $Z_i$ or $Z_iZ_j$ for the system, that indeed respect the desired fermionic symmetries as they somewhat correspond to physical operators in the encoding (hopping, number operator, and Coulomb term, respectively). We can then build a coupler by tensoring these expressions with some relevant Pauli matrix (or string) on the environment. 

\paragraph{Coulomb couplers} Similarly to the free couplers, one may consider the case where only the Coulomb term $U$ is present. We can build couplers by straightforwardly noticing that excited states are those where there are up and down fermions on the same sites, and the multiple ground states are those in which they are completely kept separate. It is then straightforward to build operators that move the fermions from one site to another. 

From our numerical simulations, it appears that both of these designs are less effective than free couplers. It might be the case that a combination of the different coupler building methods results in a better design, however. 

\begin{table}[H]
    \centering
    \begin{tabular}[t]{c|l}\hline\hline
    Coupler & Fidelity improvement\\\hline
     $V_{(1,0)}$ & $0.0031$ \\
    $V_{(2,0)}$ &  $0.0733$ \\
    $V_{(3,0)}$ &  $0.0589$ \\
    $V_{(4,0)}$ &  $0.0373$ \\
    $V_{(5,0)}$ &  $0.0000$ \\
    $V_{(6,0)}$ &  $0.0000$ \\
    $V_{(7,0)}$ &  $0.0000$ \\
    $V_{(8,0)}$ &  $0.0000$ \\
    $V_{(9,0)}$ &  $0.0000$ \\
    $V_{(10,0)}$ & $0.0000$ \\
    $V_{(11,0)}$ & $0.0000$ \\
    $V_{(12,0)}$ & $0.0000$ \\\hline\hline
\end{tabular}
\begin{tabular}[t]{c|l}\hline\hline
 Coupler & Fidelity improvement\\\hline
    $V_{(13,0)}$ & $0.1286$ \\
    $V_{(14,0)}$ & $0.3846$ \\
    $V_{(15,0)}$ & $0.0110$ \\
    $V_{(16,0)}$ & $0.3528$ \\
    $V_{(17,0)}$ & $0.4415$ \\
    $V_{(18,0)}$ & $0.2341$ \\
    $V_{(19,0)}$ & $0.1952$ \\
    $V_{(20,0)}$ & $0.2774$ \\
    $V_{(21,0)}$ & $0.0301$ \\
    $V_{(22,0)}$ & $0.0465$ \\
    $V_{(23,0)}$ & $0.3382$ \\\hline\hline
\end{tabular}
\begin{tabular}[t]{c|l}\hline\hline
     Coupler & Fidelity improvement\\\hline
    $V_{(24,0)}$ & $0.0232$ \\
    $V_{(25,0)}$ & $0.0000$ \\
    $V_{(26,0)}$ & $0.0000$ \\
    $V_{(27,0)}$ & $0.0000$ \\
    $V_{(28,0)}$ & $0.0000$ \\
    $V_{(29,0)}$ & $0.0000$ \\
    $V_{(30,0)}$ & $0.0000$ \\
    $V_{(31,0)}$ & $0.0000$ \\
    $V_{(32,0)}$ & $0.0000$ \\
    $V_{(33,0)}$ & $0.0297$ \\
    $V_{(34,0)}$ & $0.4322$ \\
    $V_{(35,0)}$ & $0.0684$ \\\hline\hline
\end{tabular}
    \caption{Fidelity improvement after a spectroscopy run using only the coupler $V_{(k,0)}$. The initial state has a fidelity of $0.4509$ with the ground state.}
    \label{tab:individualcouplers}
\end{table}

Regardless of the choice of couplers, spectroscopy needs only to be run once. Since each coupler has specific and unchanging resonances (which are not known beforehand), an analysis of the fridge energy provides the necessary information to perform the ideal version of the algorithm afterwards (with non-ideal couplers). Table \ref{tab:individualcouplers} illustrates an interesting issue: one remarks the great discrepancy in performance, as some couplers are much more critical to successful cooling than others; this, however, cannot be known in advance.

\section{Noise in the cooling process}\label{app:noisestudy}

As discussed before, one may not avoid some heating elements during the process, especially for values of $\alpha$ that stretch the RWA's applicability. Therefore, we can assume all cooling runs are noisy, with a noise channel identified with the suppressed heating process in Eq.~\ref{eq:tilde_coupler}. In experiments, when the RWA is respected, the heating channel is suppressed and cooling dominates, and one may conclude the break-even point is passed. Heating at this level is, however, suppressed by definition, so robustness to noise generated by external processes is not guaranteed in this model. 

We study two varieties of noise: totally depolarizing noise, which, during each cooling step, nudges the system density matrix towards the maximally-mixed state, and fermionic-conserving noise, which respects the symmetries of the system, and keeps the overall fermionic state in the same parity and number subspace, while nudging it towards a mixed state over all correct fermionic occupations states. 

The noise is applied after every cooling step, as the environment returns to its ground state, the system density matrix is mixed with the maximally-mixed state of the considered subspace. After the $n^{\text{th}}$ cooling step,
\begin{align}\label{eq:nudgenoise}
    \rho_{sys,n+1} &= (1-\lambda)\rho_{sys,n}+\frac{1}{\Lambda}\lambda I
\end{align}
where $\lambda \in [0,1]$, and $\Lambda$ is the normalization constant. In the case of the parity-conserving noise, we define the maximally mixed state to be
\begin{align}
    \rho_{mixed} &= \frac{1}{\Lambda}\sum_{j \in \mathcal{J}(N_f)} \ket{j}\bra{j}
\end{align}
i.e. we sum over the states that have the correct up- and down-spin quantity, $\sigma_{\uparrow}$ and $\sigma_{\downarrow}$ respectively. 

The effects of fully depolarizing noise can be avoided with the subspace cooling algorithm. It is obvious that an algorithm whose runtime is short will perform better than one whose runtime is long; it is exactly what happens here. If we perform noisy spectroscopy over the whole spectrum, many peaks appear, and the information we wish to acquire is hidden. However, if perform beforehand a pseudo-adiabatic sweep that can also be noisy, and then use the controlled cooling algorithm over the subspace, depolarization noise becomes surmountable.

\begin{figure}[ht!]
\centering
    \includegraphics{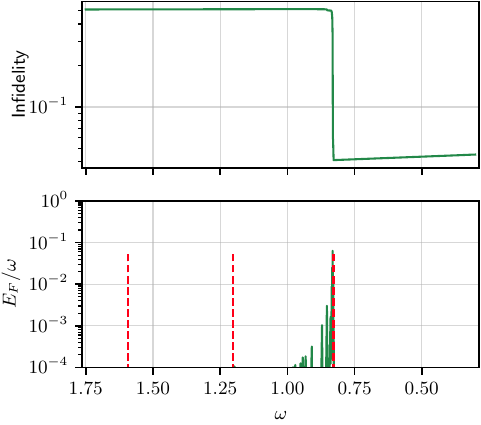}
    \caption{Subspace cooling run with parameters as in Fig.~\ref{fig:normalcontrolledrun}. Depolarizing noise is applied at each cooling step, as well as during the pseudo-adiabatic sweep, after each Trotter step, with parameter $\lambda=10^{-4}$. We start here with a Slater determinant}
    \label{fig:noisycontrolledrun}
\end{figure}

We display here noisy simulations: Fig.~\ref{fig:noisycontrolledrun} shows that subspace cooling is successful for fully depolarizing noise when $\lambda \leq 10^{-4}$. In this case, the depolarizing channel sends the state towards the maximally-mixed state. For this reason, we use the pseudo-adiabatic sweep, and cool over a restricted subspace; cooling steps need to be applied with moderation, since each wasted step is now detrimental to cooling. We end with a fidelity of $0.954$; without the pseudo-adiabatic sweep, we reach $0.877$. This is not much worse than the noiseless case in Fig.~\ref{fig:slaterfastsweepvm} (where one reaches a fidelity of $0.994$), implying that the subspace cooling algorithm can withstand higher noise. 

\begin{figure}[ht!]
\centering
    \includegraphics{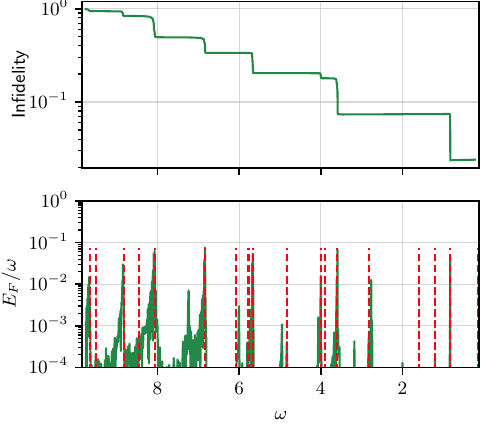}
    \caption{Noisy controlled cooling, with a maximally-mixed state restricted to the correct fermionic subspace, i.e. with $\rho = \sum_{k\in\{n_{\uparrow},n_{\downarrow}\}} \ket{E_k}\bra{E_k}$, and $\lambda=10^{-5}$, and model parameters as in Figure  \ref{fig:normalcontrolledrun}.}
    \label{fig:noisycontrolledrunsubspace}
\end{figure}

Fig.~\ref{fig:noisycontrolledrunsubspace} shows that, when we restrict the noise to only drive transitions within the correct fermionic subspace, we can scan the whole energy spectrum without any difficulty. The main difference when compared to the noiseless case is the presence of peaks on levels that are not initially populated. The action of noise diverts some of amplitude towards them, which is then picked up by the cooling algorithm. Spectroscopy informs us of supplementary peaks that need to be cooled precisely because of noise. 

Therefore, noise in the subspace increases the number of cooling steps needed, and we can maintain the ground state indefinitely as long as we cool it. In the case of fully depolarizing noise, any state eventually decays to the maximally-mixed state regardless of cooling attempts. The solution is to be fast enough to avoid it. One might also consider building a Hamiltonian from the invariants to ``cool'' back from the non-symmetry subspace to the symmetry subspace, using the same cooling algorithm. \textcite{Young_Sarovar_2013} also discuss ideas for error correction by cooling.

\subsection{Approximate noise threshold}\label{app:appnothresh}
We now derive an approximate bound for the depolarizing noise for a theoretical noisy quantum computer, above which heating surpasses cooling, and the process fails. For simplicity, let us here assume that $E_0=0$, such that the largest energy gap simplifies to $E_{max}-E_0=E_{max}$.

\setcounter{definition}{3}
\begin{theorem}\label{thm:noise}
    Let $H_j$ be a Hamiltonian as in Proposition \ref{theo:Hj_cool} governing the dynamics of cooling the $j^\text{th}$ excited state $\ket{E_j}$ down to the groundstate $\ket{E_0}$ of a system Hamiltonian $H_S$. Consider the system undergoing a global depolarizing channel with an error rate $p$. The system admits a positive net cooling rate, i.e. popoulation transfer from $\ket{E_j}$ to $\ket{E_0}$, if
    \begin{align}
        p &<  \frac{4E_j^2}{W^2 d_c^4 E_{max}^2 \pi^2},
    \end{align}
    with $W$ some constant and $d_c$ the dimension of the subspace populated by the initial state, and under the assumption that $p \ll 1$.
    \begin{proof}
        We consider heating and cooling processes as channels on the system with heating and cooling rates $\gamma_{heat}$ and $\gamma_{cool}$, respectively, determining the energies that enter and leave the system during a single cooling step, using a single coupler. It takes a wall time of $t_{comp}$ to simulate a time $t_{step}$. For this single step, the heating rate is
\begin{align}
    \gamma_{heat} = (1-(1-p)^{t_{comp}}) E_{max}/2,
\end{align}
where $E_{max}$ is the highest eigenenergy, contributing to $\gamma_{heat}$ as the mean energy $\frac{E_{max} - E_0}{2} = \frac{E_{max}}{2}$ introduced by the depolarizing channel. This approaches $E_{max}/2$ as $t_{comp}\to\infty$, telling us that the depolarizing channel pushes the system's energy towards that of the maximally mixed state as time goes on. We can break down $t_{comp}$ into the depth $D_{Trotter}$ of the Trotter circuit multiplied with a sample overhead $S_F = 1/P_j$ determined by Eq.~\eqref{eq:temp_bound}, necessary for the fridge to be excited once, 
\begin{align}
    t_{comp}=D_{Trotter} \cdot S_F. 
\end{align}
We can also obtain a more explicit expression for the depth, from \textcite{Suzuki_85}. We know a first order Trotter error follows 
\begin{align}
    \epsilon \leq \delta t \cdot t_{step} E_{max}^2,
    \label{eq:Trotter_error}
\end{align}
where $\delta t$ is the size of the time step. The depth follows $D_{Trotter} = t_{step}/\delta t$, from which we extract $D_{Trotter} \leq t_{step}^2 E^2_{max} /\epsilon$. We can allow for a Trotter error $\epsilon$ that is at the level of hardware noise $\epsilon \sim D_{Trotter} p$. Inserting this into Eq.~\eqref{eq:Trotter_error} yields $D_{Trotter} \leq t_{step} E_{max}/\sqrt{p}$. Recalling from appendix \ref{app:rwa} that $t_{step}\sim \pi/\alpha$, and in the limit $p\ll  1$, we have $\gamma_{heat} \sim p \cdot t_{comp}E_{max}/2$, such that we finally obtain
\begin{align}
    \gamma_{heat} &\leq \sqrt{p}\frac{\pi}{P_j\alpha} \frac{E_{max}}{2}
\end{align}
The cooling rate is simply $\gamma_{cool} = E_F$, where $E_F$ is the energy of the fridge after the cooling step (see section \ref{sec:ancontrolcool}): it is the energy that left the system. We can rewrite this as 
\begin{align}
    \gamma_{cool}&=E_j \cdot P_j
\end{align}
with $E_j$ the gap being cooled. In order to achieve a positive net cooling rate, we would require $\gamma_{heat} < \gamma_{cool}$. A sufficient condition for this inequality is fulfilled whenever
\begin{align}
    \sqrt{p}\frac{\pi}{P_j\alpha} \frac{E_{max}}{2} &< E_j P_j \\
    \Rightarrow p&<  \frac{4E_j^2}{W^2 d_c^4 E_{max}^2 \pi^2}
    \label{eq:p_upper_bound}
\end{align}
where we used $P_j\sim1/d_c$ (see Eq.~\eqref{eq:temp_bound}) and divided $E_j$ by a weakening factor $W$ to ensure being in the RWA regime, $\alpha \sim E_j/W$.
    \end{proof}
\end{theorem}
The energy ratio $\frac{E_j}{E_{max}}$ scales as $\mathcal{O}\left( \frac{1}{n} \right)$ in systems in which energy is an extensive quantity. Accordingly, we expect the bound to decrease with system size. Note that, in the case of thermalization, this might actually contribute to move towards the target state, although this particular calculation only concerns cooling.

\section{Supplementary cooling results}
\label{app:suppcoolresults}
 
\begin{figure}[ht!]
    \centering
    \includegraphics[width=.5\columnwidth]{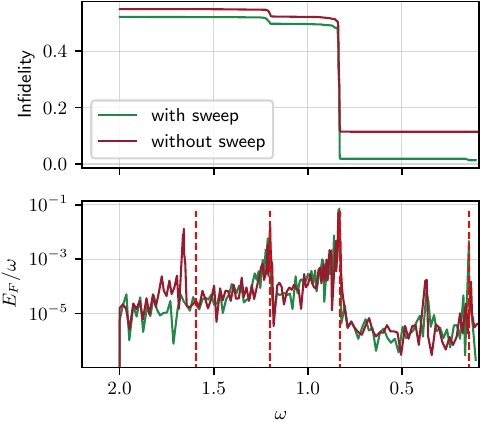}
    \caption{Effect of an initial pseudo-adiabatic sweep on a controlled cooling run. Parameters are the same as Fig.~\ref{fig:normalcontrolledrun}; we only scan, however, the lower end of the spectrum. The data in purple shows the result using the usual controlled cooling algorithm and starting in the free ground state, whereas the green lines show the improvement due to a pseudo-adiabatic sweep performed before cooling begins, from the free ground state to the interacting ground state. The time budget of the pseudo-adiabatic sweep is a hundredth of the normal sweep time.} 
    \label{fig:initadiabsweep}
\end{figure}

\begin{figure}[ht!]
    \centering
    \includegraphics[width=.5\columnwidth]{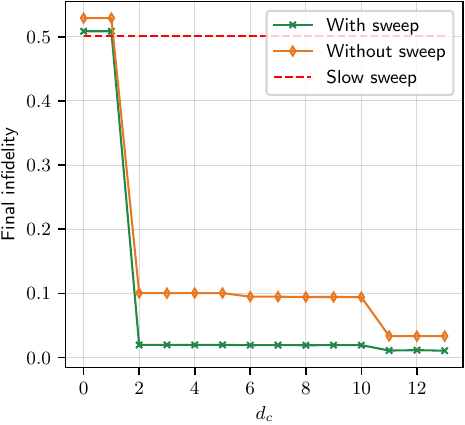}
    \caption{Counterpart to Fig.~\ref{fig:coulombfastsweepvm}. Here, the pseudo-adiabatic sweep starts from the $U=0$ ground state, instead of the $t=0$ ground state, but all other parameters are the same. The red dashed line, which shows the slow sweep, is almost unnoticeably better than the pseudo-adiabatic sweep.} 
    \label{fig:slaterfastsweepvm}
\end{figure}

We show on Fig.~\ref{fig:initadiabsweep} a full spectroscopy run with a pseudo-adiabatic sweep beforehand (not pictured). This correspond to a single point in Fig.~\ref{fig:coulombfastsweepvm} or Fig.~\ref{fig:slaterfastsweepvm}. In the upper subplot, we can see that the initial difference due to the pseudo-adiabatic sweep is barely noticeable, but the state that results is then much more amenable to cooling; this translates into the significant fidelity difference at the end of the scan. 

In Fig.~\ref{fig:slaterfastsweepvm}, we display the complementary case to Fig.~\ref{fig:coulombfastsweepvm}, in which we start from the Coulomb or $t=0$ limit. Again, the pseudo-adiabatic sweep's meager advantage translates into a larger fidelity increase after cooling. The downside in the $U=0$ case is the preparation cost of the initial ground state, which scales polynomially with the number of qubits, whereas the $t=0$ is constant, since it is a computational state. The complete adiabatic sweep performs somewhat better than the combined algorithm, but saturates at a low fidelity, since the populations spread over the degenerate subspace. 

\section{Thermal state preparation}
\label{app:thermalstateeprep}
A problem related to ground state cooling is the preparation of thermal state. In particular, we are interested in sampling from the state
\begin{align}
    \rho_S (\beta) = \frac{1}{Z_S(\beta)} e^{-\beta H_S}, \qquad Z_S(\beta) = \Tr\left( e^{-\beta H_S} \right) = \sum_j e^{-\beta E_j}
\end{align}
given the Fermi-Hubbard Hamiltonian as a system Hamiltonian as described in section \ref{app:suppcouplers}. In the limit $\beta \to \infty$, we return to the ground state problem, whereas for $\beta \to 0$ the target state is maximally mixed rendering the problem trivial. Note that, since we restrict the Hilbert space to a particle number and spin conserving subspace, the state $\rho_S$ reads
\begin{align}
    \rho_S(\beta) = \frac{1}{Z_S(\beta)} \sum_{j \in \mathcal{J}(N_f)}e^{- \beta E_j} \ket{E_j} \bra{E_j},
    \label{eq:thermal_S}
\end{align}
with $\mathcal{J}(N_f)$ indexing the symmetry subspace. Just as in section \ref{app:coolingderivation}, we start with an accessible state such as $\rho_S(0)$ and couple it to an environment, in which we can control the temperature. Sticking to a single ancilla qubit with the Hamiltonian $H_F = - \frac{1}{2} (Z_F + \mathds{1}_F)$, we can prepare the thermal state
\begin{align}
    \rho_F(\beta) = \frac{1}{Z_F(\beta)} \begin{pmatrix}
        e^{\frac{\beta \omega}{2} } & 0 \\
        0 & e^{- \frac{\beta \omega}{2} }
    \end{pmatrix} = \frac{e^{\frac{\beta \omega}{2}}}{Z_F(\beta)} \begin{pmatrix}
        1 & 0 \\
        0 & e^{- \beta \omega }
    \end{pmatrix},
    \label{eq:thermal_ancilla}
\end{align}
with $Z_F(\beta) = e^{\frac{\beta \omega}{2} } + e^{- \frac{\beta \omega}{2} }$, if we probabilistically prepare either $\ket{0}\bra{0}$ with probability $\frac{e^{\frac{\beta \omega}{2}}}{Z_F(\beta)}$ or $\ket{1}\bra{1}$ with probability $\frac{e^{ - \frac{\beta \omega}{2}}}{Z_F(\beta)}$. Let us start with a high temperature state in the system and a low temperature state in the fridge $\beta_S < \beta_F$ and couple the two systems with a two-level coupler $V = \ket{E_0}\bra{E_1} \otimes \ket{1} \bra{0} + H.c.$. To begin with, we assume the system to also be two-dimensional. The eigenstates and eigenvalues are the same as in Eq.~\eqref{eq:coupler_eigenstates} given that the coupler in on-resonant. As the thermal states are simplest in the eigenbases of the local Hamiltonians $\{\ket{E_j}\}_j$ and $\{\ket{0}, \ket{1}\}$, respectively, we write the time evolution operator $U = \exp\left( -it \left( H_S + \omega H_F + \alpha V \right) \right)$ in the tensor product basis 
\begin{align}
    U = \begin{pmatrix}
        e^{it \omega } & 0 & 0 & 0 \\
        0 & \cos(\alpha t) & -i \sin(\alpha t) & 0 \\ 
        0 & -i \sin(\alpha t) & \cos(\alpha t) & 0 \\ 
        0 & 0 & 0 & e^{-it \omega} 
    \end{pmatrix}.
    \label{eq:thermal_U}
\end{align}
Now we can calculate the evolved state. We are interested in the effect on the system only and trace out the fridge. We get
\begin{align}
    &\Tr_F \left( U (\rho_S(\beta_S) \otimes \rho_F (\beta_F)) U^\dagger \right) \nonumber \\
    &= \frac{e^{ \frac{\omega}{2} (\beta_S + \beta_F)}}{Z_S(\beta_S) Z_F(\beta_F)} \begin{pmatrix}
       1 + e^{ -\beta_F \omega } \cos^2(\alpha t) + e^{ -\beta_S \omega } \sin^2(\alpha t) & 0 \\
       0 & e^{ -\frac{\omega}{2} (\beta_S + \beta_F)} + e^{ -\beta_S \omega } \cos^2(\alpha t) + e^{ -\beta_F \omega } \sin^2(\alpha t)
    \end{pmatrix}.
    \label{eq:thermal_after_cool}
\end{align}
Comparing with the form of Eq.~\eqref{eq:thermal_ancilla}, we find that the output state in Eq.~\eqref{eq:thermal_after_cool} is again a thermal state at $t = \frac{\pi}{2 \alpha}$. In this case, the inverse temperature is given by $\beta = \beta_F$. We thus made the whole swap of excitations as in section \ref{app:coolingderivation}. The situation changes, however, if we consider the whole spectrum in the system instead of just the two lowest energies. Let us repeat the calculation in the case, where the thermal state from Eq.~\eqref{eq:thermal_S} has $d$ non-vanishing terms. If we leave the coupler unchanged, the dynamics will still be a swap of excitations within the two levels considered in Eq.~\eqref{eq:thermal_U} and a multiplication with relative phases $e^{-it(E_j \pm \frac{\omega}{2})}$ in the rest of the spectrum. We thus expect for the output state at $t= \frac{\pi}{2 \alpha}$ analogue to Eq.~\eqref{eq:thermal_after_cool}
\begin{align}
    &\Tr_F \left( U (\rho_S(\beta_S) \otimes \rho_F (\beta_F)) U^\dagger \right) = \frac{1}{Z_S(\beta_S)} \begin{pmatrix}
       \frac{e^{ \frac{\omega}{2} (\beta_S + \beta_F)} + e^{ -\beta_S \frac{\omega}{2} }}{Z_F(\beta_F)} &&&& \\
       & \frac{1 + e^{ - \frac{\omega}{2} (\beta_F - \beta_S) }}{Z_F(\beta_F)} &&& \\
       && \ddots &&\\
       &&& e^{-\beta_S E_j} & \\
       &&&& \ddots
    \end{pmatrix}.
    \label{eq:thermal_after_cool_full}
\end{align}
This is no longer a thermal state, that is we cannot identify an inverse temperature with Eq.~\eqref{eq:thermal_after_cool_full}. Instead, we can look at the ratio of the ground state population $P_0$ before and after
\begin{align}
    \frac{P_0^{\rm after}}{P_0^{\rm before}} = \frac{ e^{ \frac{\omega}{2} (\beta_S + \beta_F)} + e^{ - (\beta_S - \beta_F) \frac{\omega}{2} } }{e^{\beta_S \frac{\omega}{2} } Z_F(\beta_F)} = \frac{ 1 + e^{ - \beta_S \omega } }{ 1 + e^{ - \beta_F \omega}} > 1 \qquad \iff \qquad \beta_S < \beta_F.
\end{align}
While the above process lowers the energy in the system, we are not yet at a thermal state. It quickly becomes clear that we need to include all possible ladder operators between any two eigenenergies, including heating terms to allow for a proper thermalization. Only in the limit $\beta \to \infty$, the cooling transitions to the ground state become dominant.

\subsection{Circumventing the exponential scaling}\label{app:circumventionexpscale}
We here discuss a different approach to thermal state preparation that does not require an exponential number of transitions. In this case, the ancillary fridge is always prepared in its ground state, as opposed to the probabilistic reset. The couplers $V_{(j,k)} = \ket{E_k} \bra{E_j} \otimes \ket{1} \bra{0} + H.c.$ are chosen with probability $v_{(j,k)}$. In the end, we need the population $P_j = \frac{e^{-\beta E_j}}{Z_S(\beta)}$ for the $j^\text{th}$ energy eigenstate. The populations $P_j$ must satisfy the following set of rate equations in the perfect case
\begin{align}
    \dot P_j = \sum_k \left( v_{(k,j)} P_k - v_{(j,k)} P_j \right) \qquad \forall j,
\end{align}
where we used that the maximal population that can be transferred is proportional to the initial population, cf. Eq.~\eqref{eq:transfer}. In order to reach the desired steady state, the detailed balance condition has to hold
\begin{align}
    \sum_k v_{(k,j)} e^{-\beta E_k} = \left( \sum_k v_{(j,k)} \right) e^{-\beta E_j} \qquad \forall j.
\end{align}
One possible choice is $v_{(j,k)} = \frac{1}{d-1} \frac{ e^{-\beta E_k} }{Z_S(\beta)}$, for which the coupler towards state $\ket{E_k}$ is drawn from its thermal probability and the choice of the initial state with equal probability. The exponential factor $d$ poses a problem similar to the cooling case (cf. section \ref{app:ancillacontrolledcooling}), which can be resolved by preparing a better initial state with non-negligible population only in a low-energy sector, for instance via a preceding pseudo-adiabatic sweep (see section \ref{sec:subspacecooling}). Subroutine \ref{sub:stoctherm} is the result of these improvements with respect to the ideal version.

\begin{subroutine}[H]
    \caption{Stochastic thermalization step}\label{sub:stoctherm}
    \begin{algorithmic}[1]
        \Function{StochThermStep}{$\rho_S,\beta, \{(V_{(a,b)}, \{ \omega^{(a,b))}_\mu \}\}$}
        \State Pick random $V_k \gets V_{(\ell,m)}$ with prob. $v_{(\ell,m)}$
        \If{$\ell>m$}
        \State $\rho_F \gets \ket{0}\bra{0}$
        \ElsIf{$\ell<m$}
        \State $\rho_F \gets \ket{1}\bra{1}$
        \EndIf 
        \Comment{Couplers are always from a level to another, $\ell \neq m$}
        \For{$\omega_k \in \{\omega^{(\ell,m)}_{\mu}\}$}
            \State $\rho_S \gets\Call{CoolingStep}{\rho_S,\rho_F,V_k,\omega_k}$
        \EndFor
        \State\Return $\rho$
        \EndFunction
    \end{algorithmic}
\end{subroutine}

Depending on the temperature, the target state could already be close to a maximally mixed state, which renders the task trivial. In this case, no thermalization is needed. In the case of low temperature, the target state only has population in the low-energy sector, for which the sweep will prepare a good initial guess.

\section{Digitalization of cooling}\label{app:digitalizecooling}

While this work does not delve deep into the digitalization of the various algorithm introduced, we make a few remarks that can shine some light on the path to a fully working, gate-based version. As mentioned in section \ref{sec:algoricooli}, time evolution can be straightforwardly ported with Trotterization \cite{Childs_2021}, alongside methods for compression of gate sequences, such as methods in references \cite{Mansuroglu_2023, Mansuroglu_2023b}, as well as techniques to improve the pseudo-adiabatic sweep in the context of gate-based computing~\cite{Keever_Lubasch_2023}. This is a entire field by itself, but its solutions are independent of the problem at hand. The entire success of gate-based cooling hinges on the conversion of the couplers, which, due to the nature of fermionic systems, are non-local. As discussed in the couplers section \ref{sec:freecouplers}, the couplers being built in Fock space means encodings are straightforwardly applicable, and we may discuss on an equal footing with general fermionic dynamics, as is done for example in \textcite{Cade_2020}. Therefore, one must then focus on and perhaps may learn from the Pauli sums obtained from the encoded couplers. Indeed, the projectors considered throughout this work may hide some symmetries that the Pauli sums may reveal. This might also pave the way for better, more local couplers, although preliminary work has suggested that it is usually more effective to search for them, in the case of fermions, directly in Fock space.

\begin{figure}[ht!]
    \centering
    \includegraphics{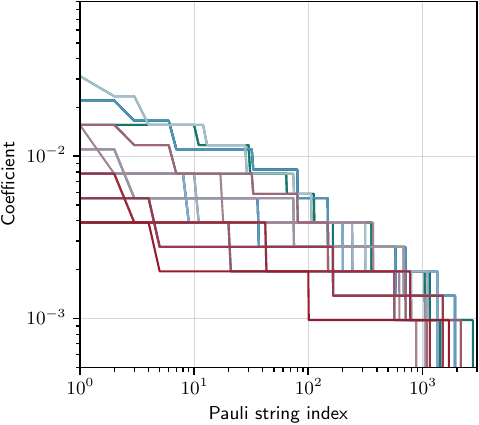}
    \caption{Sorted coefficients of the free couplers' Pauli strings. The $x$-axis represents an arbitrary Pauli string index (which is not the same for each coupler as it is sorted), and the $y$-axis the absolute amplitude of each Pauli string in the normalized sum. Each line refers to a particular coupler. In this case, we used Jordan-Wigner to encode the operators.} 
    \label{fig:freecouplersdecomposition}
\end{figure}

On Fig.~\ref{fig:freecouplersdecomposition} we show the Jordan-Wigner-encoded free couplers, in the form of the amplitudes of their Pauli strings; the dominant Pauli string for each coupler is shown in table \ref{tab:biggestpaulis}. In a future application of the subspace cooling algorithm to gate-based hardware, we would expect some truncation of the Pauli sums to be necessary: if the some of the Pauli strings are dominant, then we may start to form strategies as to which Pauli strings to keep, whereas in the opposite case, when all coefficients are equal, there is no optimal way to truncate. The free couplers' coefficients are spread over more than one order of magnitude: it seems there is hope to port the algorithm. One should also compare this table with Table~\ref{tab:individualcouplers}, which indicates that it is not necessary to run all couplers to obtain the ground state. Ideally, one would identify the optimal trade-off between coupler locality and cooling power. 

\begin{table}[H]
    \centering
    \begin{tabular}[t]{c|l}\hline\hline
    Coupler & Pauli String\\\hline
     $V_{(1,0)}$ & $X_5Z_6X_7$ \\
     $V_{(2,0)}$ & $Z_1$ \\
     $V_{(3,0)}$ & $Z_5$ \\
     $V_{(4,0)}$ & $Z_1Z_6Z_7$ \\
     $V_{(5,0)}$ & $Y_1Z_2X_3Y_5Z_6X_7$ \\
     $V_{(6,0)}$ & $X_4Z_5X_6$ \\
     $V_{(7,0)}$ & $Y_2X_3Y_6X_7$ \\
     $V_{(8,0)}$ & $Z_1X_4Z_5X_6$ \\
     $V_{(9,0)}$ & $X_2Z_4Z_5X_6$ \\
     $V_{(10,0)}$ & $Z_1X_2Z_3Z_4Z_5X_6Z_7$ \\
     $V_{(11,0)}$ & $Y_1Y_2Y_3Z_4X_5Y_6X_7$ \\
     $V_{(12,0)}$ & $Z_0$\\\hline\hline
\end{tabular}
\hfill
\begin{tabular}[t]{c|l}\hline\hline
     Coupler & Pauli String\\\hline
     $V_{(13,0)}$ & $X_3Z_4Z_5X_7$ \\
     $V_{(14,0)}$ & $Z_0Z_1$ \\
     $V_{(15,0)}$ & $Z_0Z_5$ \\
     $V_{(16,0)}$ & $Z_1Z_6Z_7$ \\
     $V_{(17,0)}$ & $Z_0Y_1Z_2X_3Y_5Z_6X_7$ \\
     $V_{(18,0)}$ & $Z_4$ \\
     $V_{(19,0)}$ & $Z_2X_3Z_4Z_5Z_6X_7$ \\
     $V_{(20,0)}$ & $Z_1Z_4$ \\
     $V_{(21,0)}$ & $Z_4Z_5$ \\
     $V_{(22,0)}$ & $Z_1Z_2Z_7$ \\
     $V_{(23,0)}$ & $Y_1Z_2X_3Z_4Y_5Z_6X_7$ \\\hline\hline
\end{tabular}
\hfill
\begin{tabular}[t]{c|l}\hline\hline
     Coupler & Pauli String\\\hline
     $V_{(24,0)}$ & $Z_0Z_6Z_7$ \\
     $V_{(25,0)}$ & $Z_0X_3Z_4Z_5X_7$ \\
     $V_{(26,0)}$ & $Z_0Z_6Z_7$ \\
     $V_{(27,0)}$ & $Z_0Z_3Z_6$ \\
     $V_{(28,0)}$ & $Z_0Z_1Z_6Z_7$ \\
     $V_{(29,0)}$ & $Z_0Z_1X_3Z_4Z_5X_7$ \\
     $V_{(30,0)}$ & $Y_0Z_1X_2Y_4Z_5X_6$ \\
     $V_{(31,0)}$ & $Y_0Z_1X_2Y_4Y_5Y_6X_7$ \\
     $V_{(32,0)}$ & $Y_0X_2Y_4Z_5X_6$ \\
     $V_{(33,0)}$ & $Y_0Z_1X_2Y_4X_6$ \\
     $V_{(34,0)}$ & $Z_0Z_1X_2Z_3Z_4Z_5X_6Z_7$ \\
     $V_{(35,0)}$ & $Y_0X_1Y_2X_3Y_4X_5Y_6X_7$ \\\hline\hline
    \end{tabular}
    \caption{Pauli string with the highest coefficient for each of the free couplers. The pauli strings in this table correspond to the index 0 coefficient in Fig.~\ref{fig:freecouplersdecomposition}.}
    \label{tab:biggestpaulis}
\end{table}

\end{document}